\documentclass[twocolumn]{aastex63}

\usepackage[utf8]{inputenc}
\DeclareUnicodeCharacter{2212}{-}
\usepackage{graphicx}
\graphicspath{{./figures/}{./}}
\usepackage{amsmath}
\usepackage{afterpage}
\usepackage{enumitem}
\setenumerate{itemsep=0mm}
\hypersetup{    
  urlcolor        = {blue},
}
\usepackage{comment}
\usepackage{multirow}

\usepackage{placeins}

\usepackage{rotating}

\usepackage{lineno}

\usepackage{soul} 
\usepackage{amsmath}
\usepackage{amssymb}
\usepackage{xspace}
\usepackage{xifthen}
\usepackage{eso-pic}

\definecolor{forestgreen}{HTML}{228B22}
\definecolor{urlblue}{HTML}{000000}

\newcommand{\FIXME}[1]{}
\newcommand{\CHECK}[1]{{#1}}
\newcommand{\response}[1]{{}}

\newcommand{\Gaia}{{\it Gaia}\xspace}
\newcommand{\SSSSS}{${S}^5$\xspace}

\mathchardef\mhyphen="2D

\newcommand{\roughly}{\ensuremath{ {\sim}\,} }

\newlength{\dhatheight}

\newcommand{\code}[1]{\texttt{#1}\xspace}

\newcommand{\unit}[1]{\ensuremath{\mathrm{\,#1}}\xspace}

\newcommand{\Myr}{\unit{Myr}}

\newcommand{\degree}{\ensuremath{{}^{\circ}}\xspace}

\newcommand{\km}{\unit{km}}
\newcommand{\kms}{\km \second^{-1}}

\newcommand{\kpc}{\unit{kpc}}
\newcommand{\second}{\unit{s}}

\newcommand{\Msun}{\unit{M_\odot}}
\newcommand{\Mstar}{\unit{M_{*}}}

\newcommand{\e}{\unit{e^{-}}}

\newcommand{\secref}[1]{Section~\ref{sec:#1}}
\newcommand{\appref}[1]{Appendix~\ref{app:#1}}
\newcommand{\tabref}[1]{Table~\ref{tab:#1}}

\newcommand{\figref}[1]{Figure~\ref{fig:#1}}

\newcommand{\bandvar}[2][]{%
  \ifthenelse{\isempty{#1}}{\var{#2}}{\var{#2\_#1}}%
}

\newcommand{\feh}{{\ensuremath{\rm [Fe/H]}}\xspace}

\newcommand{\var}[1]{\ensuremath{\texttt{\MakeUppercase{#1}}}\xspace}

\providecommand\physrep{\ref@jnl{Phys.~Rep.}}%
\providecommand\apjs{\ref@jnl{ApJS}}%
\providecommand{\jcap}{\ref@jnl{JCAP}}%

\begin{document}

\title{Measuring the Mass of the Large Magellanic Cloud with Stellar Streams Observed by \SSSSS}


\author[0000-0003-2497-091X]{Nora~Shipp}
\affiliation{Department of Astronomy and Astrophysics, University of Chicago, Chicago IL 60637, USA}
\affiliation{Kavli Institute for Cosmological Physics, University of Chicago, Chicago, IL 60637, USA}

\author[0000-0002-8448-5505]{Denis~Erkal}
\affiliation{Department of Physics, University of Surrey, Guildford GU2 7XH, UK}

\author[0000-0001-8251-933X]{Alex~Drlica-Wagner}
\affiliation{Fermi National Accelerator Laboratory, PO Box 500, Batavia, IL 60510, USA}
\affiliation{Kavli Institute for Cosmological Physics, University of Chicago, Chicago, IL 60637, USA}
\affiliation{Department of Astronomy and Astrophysics, University of Chicago, Chicago IL 60637, USA}

\author[0000-0002-9110-6163]{Ting~S.~Li}
\affiliation{Department of Astronomy and Astrophysics, University of Toronto, 50 St. George Street, Toronto ON, M5S 3H4, Canada}
\affiliation{Observatories of the Carnegie Institution for Science, 813 Santa Barbara St., Pasadena, CA 91101, USA}
\affiliation{Department of Astrophysical Sciences, Princeton University, Princeton, NJ 08544, USA}
\affiliation{NHFP Einstein Fellow}

\author[0000-0002-6021-8760]{Andrew~B.~Pace}
\affiliation{McWilliams Center for Cosmology, Carnegie Mellon University, 5000 Forbes Ave, Pittsburgh, PA 15213, USA}

\author[0000-0003-2644-135X]{Sergey~E.~Koposov}
\affiliation{Institute for Astronomy, University of Edinburgh, Royal Observatory, Blackford Hill, Edinburgh EH9 3HJ, UK}
\affiliation{Institute of Astronomy, University of Cambridge, Madingley Road, Cambridge CB3 0HA, UK}
\affiliation{Kavli Institute for Cosmology, University of Cambridge, Madingley Road, Cambridge CB3 0HA, UK}


\author[0000-0001-8536-0547]{Lara~R.~Cullinane}
\affiliation{Research School of Astronomy and Astrophysics, Australian National University, Canberra, ACT 2611, Australia}

\author[0000-0001-7019-649X]{Gary~S.~Da~Costa}
\affiliation{Research School of Astronomy and Astrophysics, Australian National University, Canberra, ACT 2611, Australia}
\affiliation{Centre of Excellence for All-Sky Astrophysics in Three Dimensions (ASTRO 3D), Australia}

\author[0000-0002-4863-8842]{Alexander~P.~Ji}
\affiliation{Observatories of the Carnegie Institution for Science, 813 Santa Barbara St., Pasadena, CA 91101, USA}
\affiliation{Department of Astronomy and Astrophysics, University of Chicago, Chicago IL 60637, USA}
\affiliation{Kavli Institute for Cosmological Physics, University of Chicago, Chicago, IL 60637, USA}

\author[0000-0003-0120-0808]{Kyler~Kuehn}
\affiliation{Lowell Observatory, 1400 W Mars Hill Rd, Flagstaff,  AZ 86001, USA}
\affiliation{Australian Astronomical Optics, Faculty of Science and Engineering, Macquarie University, Macquarie Park, NSW 2113, Australia}

\author[0000-0003-3081-9319]{Geraint~F.~Lewis}
\affiliation{Sydney Institute for Astronomy, School of Physics, A28, The University of Sydney, NSW 2006, Australia}

\author[0000-0002-6529-8093]{Dougal~Mackey}
\affiliation{Research School of Astronomy and Astrophysics, Australian National University, Canberra, ACT 2611, Australia}
\affiliation{Centre of Excellence for All-Sky Astrophysics in Three Dimensions (ASTRO 3D), Australia}

\author[0000-0002-8165-2507]{Jeffrey~D.~Simpson}
\affiliation{School of Physics, UNSW Sydney, NSW 2052, Australia}
\affiliation{Centre of Excellence for All-Sky Astrophysics in Three Dimensions (ASTRO 3D), Australia}

\author[0000-0002-3105-3821]{Zhen~Wan}
\affiliation{Sydney Institute for Astronomy, School of Physics, A28, The University of Sydney, NSW 2006, Australia}

\author[0000-0003-1124-8477]{Daniel~B.~Zucker}
\affiliation{Department of Physics \& Astronomy, Macquarie University, Sydney, NSW 2109, Australia}


\author[0000-0001-7516-4016]{Joss~Bland-Hawthorn}
\affiliation{Sydney Institute for Astronomy, School of Physics, A28, The University of Sydney, NSW 2006, Australia}
\affiliation{Centre of Excellence for All-Sky Astrophysics in Three Dimensions (ASTRO 3D), Australia}

\author[0000-0001-6957-1627]{Peter~S.~Ferguson}
\affiliation{George P. and Cynthia Woods Mitchell Institute for Fundamental Physics and Astronomy, Texas A\&M University, College Station, TX 77843, USA}
\affiliation{Department of Physics and Astronomy, Texas A\&M University, College Station, TX 77843, USA}

\author[0000-0001-9046-691X]{Sophia~Lilleengen}
\affiliation{Department of Physics, University of Surrey, Guildford GU2 7XH, UK}

\collaboration{18}{(\SSSSS Collaboration)}

\email{norashipp@uchicago.edu}

\begin{abstract}
Stellar streams are excellent probes of the underlying gravitational potential in which they evolve. In this work, we fit dynamical models to five streams in the Southern Galactic hemisphere, combining observations from the Southern Stellar Stream Spectroscopic Survey (\SSSSS), \Gaia EDR3, and the Dark Energy Survey (DES), to measure the mass of the Large Magellanic Cloud (LMC). With an ensemble of streams, we find a mass of the LMC ranging from $\sim 14$ to $19 \times 10^{10} \Msun$, probed over a range of closest approach times and distances. With the most constraining stream (Orphan-Chenab), we measure an LMC mass of $18.8^{+ 3.5}_{- 4.0} \times 10^{10} \Msun$, probed at a closest approach time of $310 \Myr$ and a closest approach distance of $25.4 \kpc$. 
This mass is compatible with previous measurements, showing that a consistent picture is emerging of the LMC's influence on structures in the Milky Way. Using this sample of streams, we find that the LMC's effect depends on the relative orientation of the stream and LMC at their point of closest approach. To better understand this, we present a simple model based on the impulse approximation and we show that the LMC's effect depends both on the magnitude of the velocity kick imparted to the stream and the direction of this kick. 
\end{abstract}


\section{Introduction}
\label{sec:intro}

The mass of the Large Magellanic Cloud (LMC), the Milky Way's largest satellite galaxy, has proven notoriously difficult to measure.
Efforts to directly measure the mass of the LMC from the dynamics of its star clusters \citep{Schommer:1992} and rotation curve \citep{vanderMarel:2014} both yield relatively modest estimates of $\roughly 2 \times 10^{10} \Msun$ within $\roughly 9 \kpc$. \CHECK{More recent measurements with \Gaia yield similar results \citep[e.g.,][]{Vasiliev2018,Cullinane:2020,Wan:2020b}.}
However, several distinct lines of reasoning suggest that the LMC may have a total mass that is up to an order of magnitude larger.
First, the LMC's large speed relative to the Milky Way \citep[e.g.,][]{Kallyvalail:2006a} is consistent with it being on its first passage around the Milky Way \citep{Besla:2007}. Given this first passage scenario and the close association of the LMC and the Small Magellanic Cloud (SMC) in both position and velocity, it is reasonable to assume that they were accreted onto the Milky Way together. In order for the SMC to have initially been gravitationally bound to the LMC, the mass of the LMC must be greater than $\roughly 10^{11}\Msun$ \citep{kallivayalil_etal_2013}. 
Second, accounting for the LMC's effect on the ``timing argument'' for the Milky Way and M31, as well as the nearby Hubble flow, gives an LMC mass of $2.5 \times 10^{11} \Msun$ \citep{Penarrubia:2016}. 
Third, $N$-body simulations of the LMC on a first-infall orbit favor a massive LMC up to $2.5 \times 10^{11} \Msun$ to explain the warp in the Milky Way's HI disk \citep{Weinberg:1998,Weinberg:2006,Levine:2006,Laporte2018}. 
Finally, abundance matching based on the stellar mass of the LMC \citep[$\Mstar = 2.7 \times 10^{9} \Msun$;][]{van_der_marel:2002} gives a peak halo mass of $\roughly 2 \times 10^{11} \Msun$ \citep{Boylan-Kolchin:2010, Moster:2013, Behroozi:2013a, Dooley:2017a, Dooley:2017b}. 
These arguments suggest that direct dynamical tracers are only measuring the central region of a much more massive LMC halo.

Stellar streams, the remnants of recently disrupted dwarf galaxies and globular clusters, provide a direct dynamical tracer of the mass of the LMC at much larger distances. 
The influence of the LMC on the behavior of stellar streams around the Milky Way was first considered in detail by \citet{Law:2010}, who discussed the interaction of a relatively light LMC ($< 6 \times 10^{10} \Msun$) with the Sagittarius stream and found that it could
have a significant effect. 
Following the same argument, \citet{Vera-Ciro:2013} showed that an LMC with a mass of $8 \times 10^{10} \Msun$ could change the shape of the Milky Way halo inferred by \citet{Law:2010}, making it more spherical. 
Along these lines, \citet{Gomez:2015} found that the infall of a $1.8 \times 10^{11}\Msun$ LMC would induce a significant reflex motion in the Milky Way, which would affect the Sagittarius stream. 

Recently, the Dark Energy Survey (DES) discovered a large number of stellar streams in the southern hemisphere \citep{Shipp:2018}. \Gaia then provided unprecedented measurements of proper motions of greater than 1 billion Milky Way stars, enabling the measurement of the proper motions of the DES streams \citep{Shipp:2019}.
Many of these streams are close in projection to the  LMC, suggesting the exciting opportunity to probe the mass of the LMC at large radii with multiple direct dynamical tracers.
Such a measurement was proposed by \citet{Erkal:2018a}, who predicted the effect of the LMC on the Tucana III (Tuc III) stream and found that
the LMC could induce a substantial proper motion perpendicular to the track of the stream on the sky. 
They further argued that the size of this offset could be used to measure the mass of the LMC.
Interestingly, the proper motion offset predicted by \citet{Erkal:2018a} was not observed by \citet{Shipp:2019}, using data from \Gaia DR2 \citep{Gaia:2018}.

The next attempt to perform this measurement came when \citet{Koposov:2019} used data from \Gaia DR2 to determine that the Orphan stream discovered in SDSS \citep{Grillmair:2006,Belokurov:2006} and the Chenab stream discovered in DES \citep{Shipp:2018} likely originated from the same progenitor.
\citet{Erkal:2019} proposed that the large-scale wobble of the joint Orphan--Chenab stream track, together with the misalignment between the track and proper motion reported by \citet{Koposov:2019}, could be best explained as a result of an interaction between the stream and the LMC.
Furthermore, \citet{Erkal:2019} were able to fit the track of the Orphan--Chenab stream in an aspherical Milky Way potential, including an infalling LMC, to simultaneously measure an LMC mass of $1.38^{+0.27}_{−0.24} \times 10^{11} \Msun$ and a Milky Way mass of $3.80^{+0.14}_{−0.11} \times 10^{11} \Msun$ within $50 \kpc$. Subsequently, \cite{Vasiliev:2021} used the Sagittarius stream to simultaneously fit the LMC and Milky Way potential and obtained an LMC mass of $(1.3\pm0.3)\times10^{11} \Msun$.

In this paper, we extend the analyses of \citet{Erkal:2018a} and \citet{Erkal:2019} to five of the DES streams with proper motions measured by \Gaia EDR3 \citep{Gaia:2016, Gaia:2021} and radial velocities measured by the Southern Stellar Stream Spectroscopic Survey \citep[\SSSSS;][]{Li:2019}. In Section \ref{sec:data}, we describe the data used in this work. In Section \ref{sec:method}, we explain how we fit each stream. In Section \ref{sec:results}, we present our measurement of the LMC mass from each stream. In Section \ref{sec:disc}, we discuss the implications of our results before concluding in Section \ref{sec:conc}.

\section{Data}
\label{sec:data}

\subsection{Observations}

The precise modeling of stellar streams requires 6D phase-space measurements, consisting of 3D positions and 3D velocities. Until recently, such measurements were available for only a small number of streams \citep[e.g.,][]{Majewski:2004, Koposov:2010, Sesar:2015, Ibata:2016}. \SSSSS, in conjunction with astrometric data from \Gaia and photometry from DES, has provided unprecedented systematic 6D measurements of over 20 streams in the Southern Hemisphere. Here we provide a brief overview of \SSSSS, and we refer readers to \citet{Li:2019} for more details.  \SSSSS uses the Two-degree Field (2dF) fiber positioner \citep{Lewis:2002} coupled with the dual-arm AAOmega spectrograph \citep{Sharp:2006} on the 3.9-m Anglo-Australian Telescope (AAT). 2dF provides 392 science fibres that can be distributed across a field of view of $\roughly 3 \deg^2$. The gratings employed were 580V on the blue arm and 1700D on the red arm, corresponding to spectral resolutions of $\sim1300$ and $\sim 10000$, respectively. The gratings were chosen to achieve the highest spectral resolution in the red centred on the near-infrared calcium triplet (CaT) lines in order to derive precise radial velocities of stream members. Both radial velocities (RVs) and stellar atmospheric parameters of each star were derived simultaneously using the \code{rvspecfit}\footnote{\url{https://github.com/segasai/rvspecfit}} template fitting code \citep{Li:2019,Koposov2019rvspecfit}. For each stream field, the average exposure time is about 2 hrs to reach S/N $\sim 5$ at $r=18.5$ for RV precision $\sim 1\,\kms$.

\subsection{Stream Selection}

In this study, we consider the southern streams which were observed with AAT by the end of 2018. ATLAS, Chenab, Elqui, Indus, Phoenix, Jhelum, and Aliqa Uma were observed by \SSSSS in 2018 \citep{Li:2019}, while the Tuc III stream was observed with the same setup as an \SSSSS pilot program \citep{Li:2018} prior to 2018. We selected five of these streams -- ATLAS, Chenab, Elqui, Indus and Phoenix, with which to fit the mass of the LMC. These streams were selected because they have complete 6D phase-space measurements, and they show no signs of significant perturbation beyond the LMC that would require additional model complexity.

We exclude three streams from our final LMC mass analysis due to evidence of clear additional perturbations.
First, we exclude Aliqa Uma, which has been shown to be an extension of the ATLAS stream \citep{Li:2021}. Aliqa Uma has been separated from the ATLAS stream by an unknown perturber. 
Since our fits only include the potential of the Milky Way and the LMC, they are unable to reproduce such small-scale features. We also examine the  streams observed by \SSSSS for evidence of perturbation by the Milky Way bar. We do this by including an analytic bar potential, as described in Section 5.2.1 of \citet{Li:2021}. For each stream, we sample the stream orbital parameters and the bar pattern speed 100 times. For the pattern speed, we use $\Omega = 41 \pm 3 \kms \kpc$ from \citet{Sanders:2019}. We compare the resulting stream models by eye to the models excluding the bar, and find that of the \SSSSS streams, only Tuc III is likely to have been significantly perturbed by the bar. We therefore exclude the Tuc III stream from our analysis. We also exclude the Jhelum stream due to evidence of perturbation reported by \citet{Bonaca:2019} and \citet{Shipp:2019}. While we exclude these streams from the LMC mass fitting analysis, we do fit models to Tuc III and Jhelum including an LMC mass fixed to $1.5 \times 10^{11} \Msun$ in order to examine the possible effect of the LMC on these streams \citep[as in][]{Wan:2020,Li:2021}.

In order to dynamically model the streams, a clean sample of spectroscopic member stars is needed for the fit. We take the spectroscopic member stars reported in \citet{Li:2021} for ATLAS, \citet{Wan:2020} for Phoenix, and \citet{Li:2018} for Tuc III. These three streams all have narrow spatial widths, small velocity dispersions, and unresolved metallicity dispersions; their progenitors are likely to be globular clusters or very low-luminosity dwarf galaxies. The stream membership is usually unambiguous, and member stars are selected subjectively based on their radial velocities, proper motions, metallicities, and locations on the color-magnitude diagram in the reference therein. 

For the Jhelum, Indus, and Elqui streams that have a large stream width and whose progenitors are likely to be classical dwarf galaxies \citep{Ji:2020}, the membership is less obvious since the streams are embedded in foreground contamination. The stream members are therefore determined with a mixture model including several multivariate Gaussian components in proper motion, radial velocity, and metallicity space, detailed in Pace et al.\ (in prep.). We selected highly probable members with membership probability $P_\mathrm{mem} > 0.8$ from the mixture models as the stream members for this work. Finally, we consider the Chenab stream. Chenab was identified as the southern extension of the Orphan stream by \citet{Koposov:2019}. \SSSSS mapped the entire Orphan--Chenab stream within the DES footprint in 2018, and partially observed some of the northern extension in 2019 \citep{Li:2019}. Through the rest of this work, we abbreviate the name of the Orphan--Chenab stream as OC.
We use a dataset that combines the \SSSSS data with data from \CHECK{LAMOST \citep{Zhao:2012} and APOGEE \citep{Majewski:2017}} to cover more than 100 degrees along the OC stream. \CHECK{Rather than fitting individual member stars, we use splines fit to this dataset by Koposov et al.\ (in prep.), similar to those of \citet{Koposov:2019} for the northern component of the stream.} Stream member stars are selected from \SSSSS works that are published or in preparation.\footnote{Note that since the \SSSSS catalog is continuously updated with new observations and with improvements to the reduction pipeline, the adopted members may change before publication of the final member list in the associated work. However, we do not expect these small differences to change our results.} We use the RA, Dec, and proper motion measurements from \Gaia EDR3 and the RVs from \SSSSS for these individual member stars as input for the modeling. Specifically, the RVs are taken from the second internal data release (iDR2.2) which is described in detail in \citet{Ji:2021}.
We note that, as all the streams presented here except for Tucana III were observed in 2018 by \SSSSS, the RVs of the these stream members are also available in the first public release (DR1)\footnote{\url{https://zenodo.org/record/4695135\#.YNv\_PjZKjdc}} of \SSSSS,  which is based on an earlier internal data release \citep[iDR1.4;][]{Li:2019} with the same spectral data. 

\subsection{Distance Measurements}

Distance is an important component of 6D phase-space measurements. Although \citet{Shipp:2018} measured the distance for all these streams with isochrone fitting, the distance gradients along the streams are much harder to constrain. In this work, we include the distance measurements from individual blue horizontal branch stars (BHBs) and RR Lyrae stars (RRLs) following the same methods described in \citet{Li:2021}. In short, we cross match the spectroscopic stream members with the \Gaia RRL catalogues \citep{Clementini2018,Holl2018} to find the RRL members in these streams. We then determine the distance modulus of RRL stars using the relation from \citet{Muraveva:2018} and dereddened \Gaia G-band magnitude. The relation is metallicity dependent; we therefore adopt the mean metallicity of each stream from the literature for the distance determination.
We classify stream members with $g-r < 0.1$ and not in RRL catalogues as BHB members. 
The distance modulus of each BHB is then calculated using the relation from \citet{Belokurov:2016} and dereddened DES photometry. 
In our current sample, we do not see a systematic offset in distance between the two populations of tracers.
We assume an uncertainty in distance modulus of 0.17 mag from \citet{Muraveva:2018} for RRLs, and an uncertainty of 0.1 mag from \cite{Deason:2011} for BHBs.  We note that we only selected BHB and RRL members from the spectroscopic sample. Although more BHB and RRL members are likely present outside of AAT fields, especially for the dwarf galaxy streams with large stream widths, we limit our selection to the stars that have radial velocities available for a purer sample. We list the number of spectroscopic members along with the number of BHBs and RRL members used in each stream in \secref{individual} when we discuss individual streams.

\section{Method}
\label{sec:method}

\subsection{Stream Models}

Following the method of \citet{Erkal:2019}, we model stream formation and evolution using the modified Lagrange Cloud Stripping (mLCS) technique developed in \citet{Gibbons:2014}. The method consists of releasing test particles at the Lagrange points of the progenitor, and then evolving them in the combined potential of the the progenitor, the Milky Way and the LMC. As in \citet{Erkal:2019}, we model both the Milky Way and LMC as individual particles sourcing their respective gravitational potentials, which is crucial for capturing the response of the Milky Way to the LMC. 

We represent the Milky Way potential using the results of \citet{McMillan:2017}, and evaluate the acceleration from the potential using \code{galpot} \citep{Dehnen:1998}. The \citet{McMillan:2017} potential includes six axisymmetric components, namely bulge, dark matter halo, thin and thick stellar disk, and HI and molecular gas disks. We take the Sun's position and 3D velocity from \citet{McMillan:2017}. As described in \citet{Li:2021} and \citet{Wan:2020}, in order to examine the effects of the choice of Milky Way potential model on our stream fits, we produce ten realizations of this potential by sampling the Markov Chain Monte Carlo (MCMC) chains from the fit in \citet{McMillan:2017}. We find that the least massive of these realizations ($M_{\rm MW} = 8.3 \times 10^{11} \Msun$) provides the best fit to the stream data. We provide the potential parameters in Table~\ref{tab:potential} in the same format as \citet{McMillan:2017}.

We model the mass distribution of the LMC as a stellar disk and a dark matter halo. The stellar disk is modeled as a Miyamoto-Nagai disk \citep{Miyamoto:1975} with a mass of $3\times10^9 \Msun$, a scale radius of $1.5$ kpc, and a scale height of $0.3$ kpc. The orientation of the LMC disk matches the measurement of \cite{vanderMarel:2014}. The LMC's dark matter halo is modeled as a Hernquist profile \citep{Hernquist:1990}. As in \citet{Erkal:2019}, we leave the total mass of the LMC as a free parameter but fix the scale radius to match the circular velocity measurement of $91.7 \kms$ at $8.7 \kpc$ from \cite{vanderMarel:2014}. Note that this is in agreement with more recent measurements of the LMC's circular velocity \citep[e.g.,][]{Cullinane:2020}. The total mass and shape of the LMC potential is fixed throughout each simulation, and does not evolve after infall. We account for the dynamical friction of the Milky Way on the LMC using the results of \cite{Jethwa:2016}. We also fit for the proper motion, distance, and radial velocity of the LMC with priors given by their observed value and uncertainty \citep{kallivayalil_etal_2013,Pietrzyski:2013,van_der_marel:2002}.

We model the potential of each stream's progenitor as a Plummer sphere \citep{Plummer:1911} with a mass and scale radius chosen to match the observed stream width. During the course of tidal disruption, the progenitor's mass decreases linearly in time to account for tidal stripping. The majority of the streams considered in this work do not have a known progenitor, so we assume that the progenitor has completely disrupted, i.e., that its present day mass is zero. Furthermore, for the majority of streams, we assume that the remnant of the progenitor is in the middle of each stream's observed extent. The only exceptions are Tuc III and OC. Tuc III is one of the few streams known to be associated with a bound progenitor. For this stream, we require a bound progenitor to remain at present day, positioned at $(\phi_1, \phi_2) = (0 \degr, 0 \degr)$ in the stream coordinates of Tuc III. For OC, the progenitor is placed at $\phi_1 = 6.34 \degr$ in the coordinate system from \citet{Erkal:2019} and \citet{Koposov:2019}. This is near the center of the full OC stream, and is the same progenitor position as in \citet{Erkal:2019}. For all streams we use coordinate systems defined by the rotation matrices in Appendix D of \citet{Shipp:2019}, with the exception of OC, for which we use the coordinate system defined by \citet{Koposov:2019}.

\subsection{Comparison with Data}

We calculate the likelihood of each stream model by producing mock observations of the simulated stream and comparing them with the data set described above. For each stream model, we calculate the track on the sky, the radial velocity, the proper motions in RA and Dec, and the distance as functions of $\phi_1$, the observed angle along the stream. The likelihood is calculated for each \SSSSS member star, as described in Section 3.2 of \citet{Erkal:2019}, using simulated particles within $\pm 1\degree$ in $\phi_{1}$ of each \SSSSS member. When calculating the likelihood of the stream track and radial velocity of each star, we take into account not only the measurement uncertainty of each star, but also the intrinsic width and stellar velocity dispersion. We introduce two nuisance parameters ($\sigma_{\rm \phi_2}$, $\sigma_{\rm vr}$), which are added in quadrature to the intrinsic width and velocity dispersions of the model.
This allows for slight variations in the stream model to better fit the observed data without varying the progenitor mass. We assign the total mass of each progenitor in order to roughly reproduce the observed width of each stream by eye. The progenitor parameters are listed in \tabref{params}.

We perform a MCMC fit using \code{emcee} \citep{Foreman_Mackey:2013}. Our model includes 12 free parameters, namely the $\phi_2$ position, distance, radial velocity, and proper motion of the progenitor at present day, the stream track and radial velocity nuisance parameters, and the proper motion, radial velocity, distance, and total mass of the LMC.  
The prior distributions on each parameter are listed in \tabref{priors}. \CHECK{The upper limit of our prior on the LMC mass is selected to ensure that the LMC is on its first infall within our Milky Way potential. The priors on the stream progenitor properties are selected to be uninformative.}

\begin{figure*}[htb!]
\centering
\includegraphics[width=0.7\textwidth]{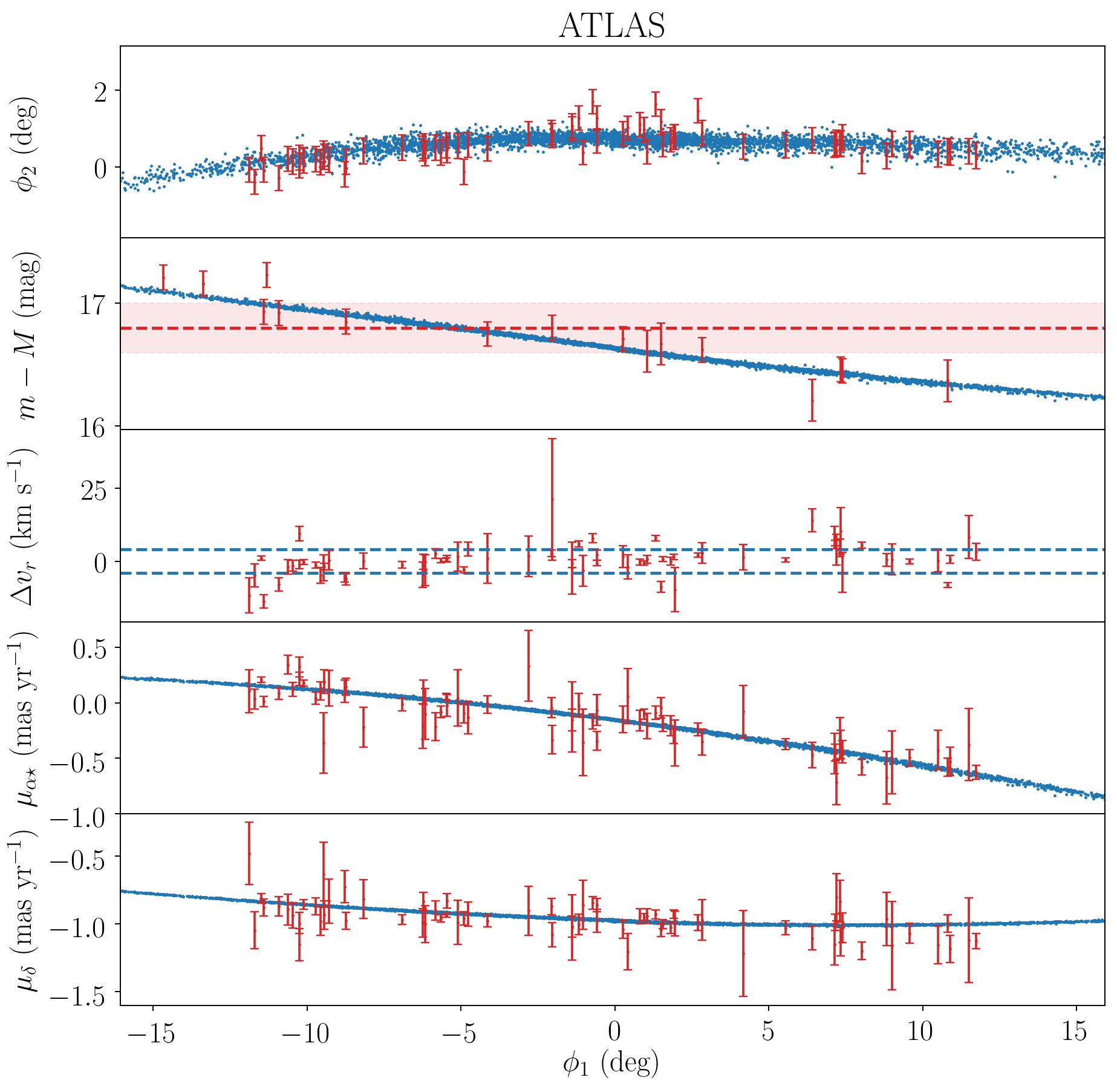}
\caption{Model fit to the ATLAS stream. The simulated stream is shown in blue, and the \SSSSS members included in the MCMC fit are plotted in red. In the first panel, we present the track in stream coordinates ($\phi_1, \phi_2$), calculated using the rotation matrices from \citet{Shipp:2019} and the RA and Dec measurements from \Gaia EDR3. The error bars on the data represent the total stream width fit to the data.
In the second panel, we show the distance modulus along the stream. The dashed red line and shaded region represent the distance measurement from \citet{Shipp:2018}  with a 0.2 mag uncertainty, and the individual points represent the BHB and  RRL  distance tracers included in the fit. In the third panel, we plot the difference between the measured and model radial velocity at the $\phi_1$ position of each member star. The separation of the blue dashed lines is equal to two times the total velocity dispersion fit to the data. 
The fourth and fifth panels show the proper motions of the model and the measured proper motions of the \SSSSS member stars from the \Gaia EDR3 dataset. Similar figures for each stream are included in \appref{models}.}
\label{fig:atlas}
\end{figure*}

\newcommand{\priorcaption}{Priors on MCMC fit parameters.}
\newcommand{\priorcomments}{Priors on the twelve free parameters in the MCMC stream model fits. $m - M_{\rm measured}$ is the distance modulus reported in Table 1 of \citet{Shipp:2018}.}
\begin{deluxetable*}{l ccccc}
\tablecolumns{13}
\tablewidth{0pt}
\tabletypesize{\scriptsize}
\tablecaption{ \priorcaption }
\label{tab:priors}
\tablehead{
\colhead{Parameter} & \colhead{Prior} & \colhead{Range} & \colhead{Units} & \colhead{Description} }
\startdata
$\phi_{\rm 2, prog}$           & Uniform     & (-1, 1)           & deg                    & Vertical spatial placement of the progenitor in stream coordinates. \\
$\sigma_{\rm \phi_2, prog}$    & Uniform     & (0, 2)            & deg                    & Nuisance parameter representing stream spatial width. \\
$v_{\rm r, prog}$              & Uniform     & (-300, 300)       & $\kms$                    & Radial velocity of the progenitor. \\
$\sigma_{\rm v_r, prog}$       & Uniform     & (0, 20)           & $\kms$                    & Nuisance parameter representing stream velocity dispersion. \\
$(m-M)_{\rm prog}$             & Normal      & $(m-M)_0 \pm 0.2$ & mag                    & Distance modulus of the progenitor. \\
$\mu_{\rm \phi_1 \star, prog}$ & Uniform     & (-10, 10)         & $\mathrm{mas\ yr^{-1}}$ & Proper motion of the progenitor along the stream. \\
$\mu_{\rm \phi_2, prog}$       & Uniform     & (-10, 10)         & $\mathrm{mas\ yr^{-1}}$ & Proper motion of the progenitor perpendicular to the stream. \\
\tableline
$M_{\rm LMC}$                  & Log-Uniform & (2, 30)           & $10^{10} \Msun$         & Total mass of the LMC. \\ 
$\mu_{\rm \alpha \star, LMC}$  & Normal      & $1.91 \pm 0.02$   & $\mathrm{mas\ yr^{-1}}$ & Proper motion of the LMC in RA \citep{kallivayalil_etal_2013}. \\
$\mu_{\rm \delta, LMC}$        & Normal      & $0.229 \pm 0.047$ & $\mathrm{mas\ yr^{-1}}$ & Proper motion of the LMC in Dec \citep{kallivayalil_etal_2013}. \\
$v_{\rm r, LMC}$               & Normal      & $262.2 \pm 3.4$   & $\kms$                    & Radial velocity of the LMC \citep{van_der_marel:2002}. \\
$d_{\rm LMC}$                  & Normal      & $49.97 \pm 1.13$  & \kpc                    & Distance of the LMC \citep{Pietrzyski:2013}. \\
\tableline
$M_{\rm prog}$                 & Fixed       & --                & \Msun                   & Mass of the progenitor (see Tab.~\ref{tab:params}). \\
$r_{\rm s, prog}$              & Fixed       & --                & \kpc                    & Scale radius of the progenitor (see Tab.~\ref{tab:params}).
\enddata
\end{deluxetable*}


\section{Results}
\label{sec:results}

\newcommand{\resultscaption}{LMC mass measurements and parameters of the last closest approach.}
\begin{deluxetable}{l cccc}
\tablecolumns{13}
\tablewidth{0pt}
\tabletypesize{\scriptsize}
\tablecaption{ \resultscaption }
\label{tab:results}
\tablehead{\colhead{Stream} & \colhead{$M_{\rm LMC}$} & \colhead{$r_{\rm approach}$} & \colhead{$v_{\rm approach}$} & \colhead{$t_{\rm approach}$} \\[-0.5em]
 & $(10^{10}\ \Msun)$ & (\kpc) & ($\kms$) & (\Myr) 
}
\startdata
ATLAS & $14.3^{+ 6.7}_{- 3.5}$ & 23.9 & 467.1 & 80.0 \\
OC & $18.8^{+ 3.5}_{- 4.0}$ & \CHECK{25.4} & \CHECK{371.2} & \CHECK{310.0} \\
Elqui & $16.8^{+ 5.2}_{- 3.0}$ & 11.2 & 419.6 & 99.0 \\
Indus & $15.6^{+ 8.6}_{- 3.6}$ & 38.0 & 268.5 & 10.5 \\
Phoenix & $2.7^{+ 8.5}_{- 0.7}$ & 30.7 & 433.9 & 49.2 \\
Tucana III & - & 4.2 & 382.4 & 98.8 \\
Jhelum & - & 40.6 & 367.2 & 2.8 
\enddata
\end{deluxetable}

\begin{figure*}[htb!]
\centering
\includegraphics[width=0.95\textwidth]{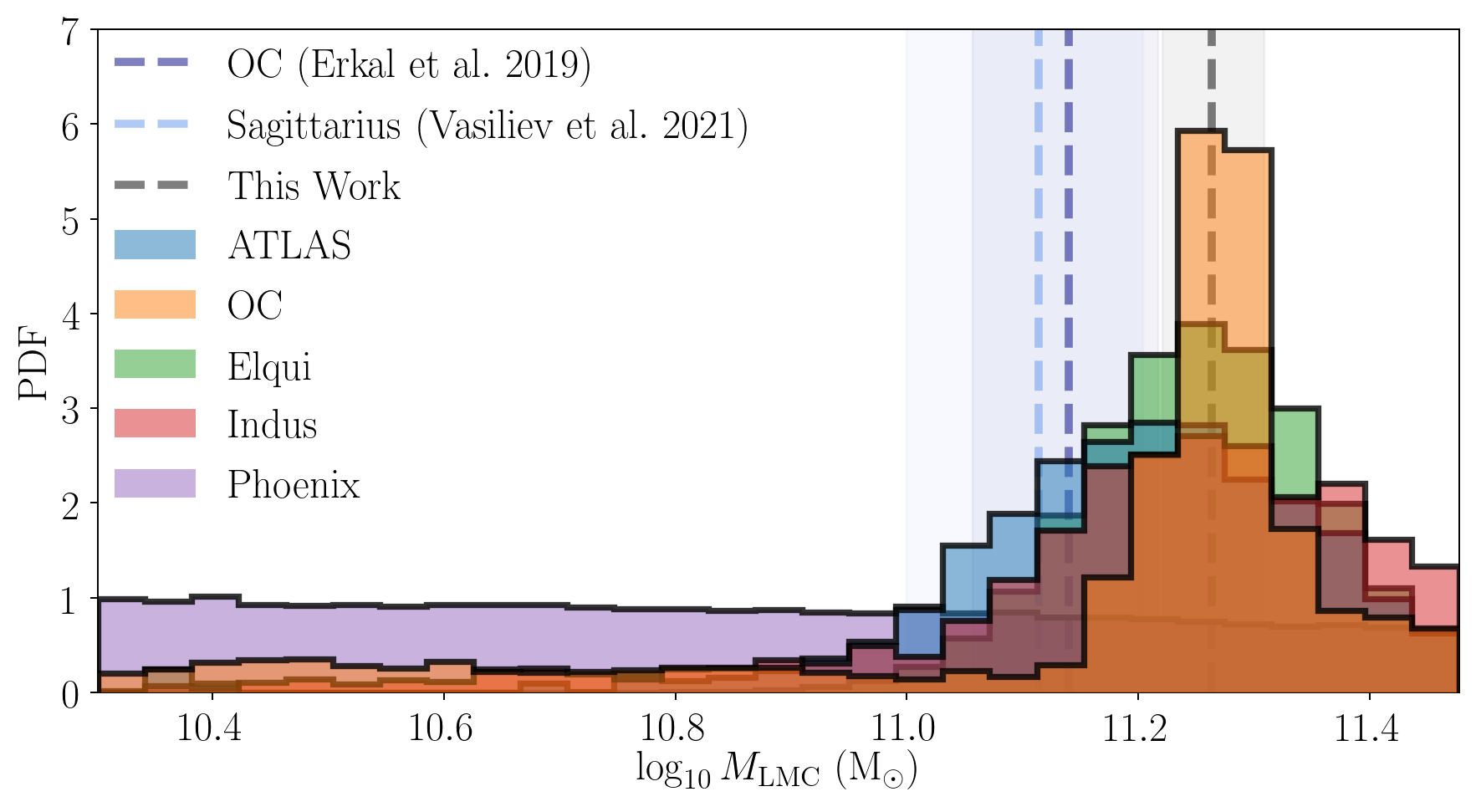}
\caption{Marginalized posterior distributions on the total LMC mass from fits to each of the five streams. The dark blue vertical dashed line and shaded region show the constraint on the LMC mass from the fit to the OC stream in \citet{Erkal:2019}, the light blue dashed line and shaded region represent the measurement using the Sagittarius stream by \citet{Vasiliev:2021}, and the gray dashed line and shaded region represent the combined mass constraint from the five measurements presented in this work.}
\label{fig:lmc_mass}
\end{figure*}

Following the method described above, we fit models to each stream and obtain five independent constraints on the total mass of the LMC. As a demonstration of the stream models in comparison to the data, we show the best-fit model to the ATLAS stream in Figure \ref{fig:atlas}. The red points are the \SSSSS members and the blue points represent the stream model.  This shows that the best-fit model can recover all of the observed trends in the ATLAS stream. Similar figures for the other streams are included in \appref{models}. \CHECK{In addition,  a movie showing the orbits of the seven streams included in this work and the LMC can be found at \href{https://www.youtube.com/watch?v=CrYhjJ-u5RA}{this link}.}\footnote{\url{https://www.youtube.com/watch?v=CrYhjJ-u5RA}}

In this section, we present the measurements of the mass of the LMC, and discuss the details of the interaction between each stream and the LMC in order to develop a consistent picture of how the Milky Way's largest satellite has perturbed this population of stellar streams.

The constraints on the LMC mass are presented in Table \ref{tab:results} and Figure \ref{fig:lmc_mass}. Constraints on the other progenitor and LMC parameters are included in \tabref{params}. In Figure \ref{fig:lmc_mass}, each color represents the posterior distribution on the LMC mass resulting from the fit to each stream and marginalized over the other fit parameters. The dark blue shaded region represents the best-fit LMC mass and uncertainty from an analysis of the OC stream \citep{Erkal:2019}, and the light blue shaded region represents LMC mass inferred from fits to the Sagittarius stream \citep{Vasiliev:2021}. Generally, the results are consistent with each other and the two previous measurements to within $1 \sigma$. \CHECK{We thus combine the individual measurements to provide a joint constraint on the LMC mass, which is displayed as the gray shaded region and described in \secref{combined}.} We note that the posterior distribution for the Phoenix stream is very broad and provides no meaningful constraint on the LMC mass. 

\subsection{Physical Intuition}

Stream perturbations can manifest as a difference between the spatial and velocity components. This can be observed as a misalignment between the proper motion ($\mu_{\rm \phi_2} / \mu_{\rm \phi_1}$) and stream track ($ d \phi_2 / d \phi_1$), or as an offset between the radial velocity ($v_{\rm r} / \mu_{\rm \phi_1}$) and distance ($d D / d \phi_1$) gradients along the stream (see Sec.~\ref{sec:offsets} for more details). The current data are most sensitive to the on-sky perturbation (i.e., offsets between the proper motion and stream track), because distance gradients measured along the stream are relatively imprecise. Thus, our constraints on the LMC mass depend both on the strength and direction of the LMC perturbation.

\begin{figure*}[htb!]
\centering
\includegraphics[width=0.8\textwidth]{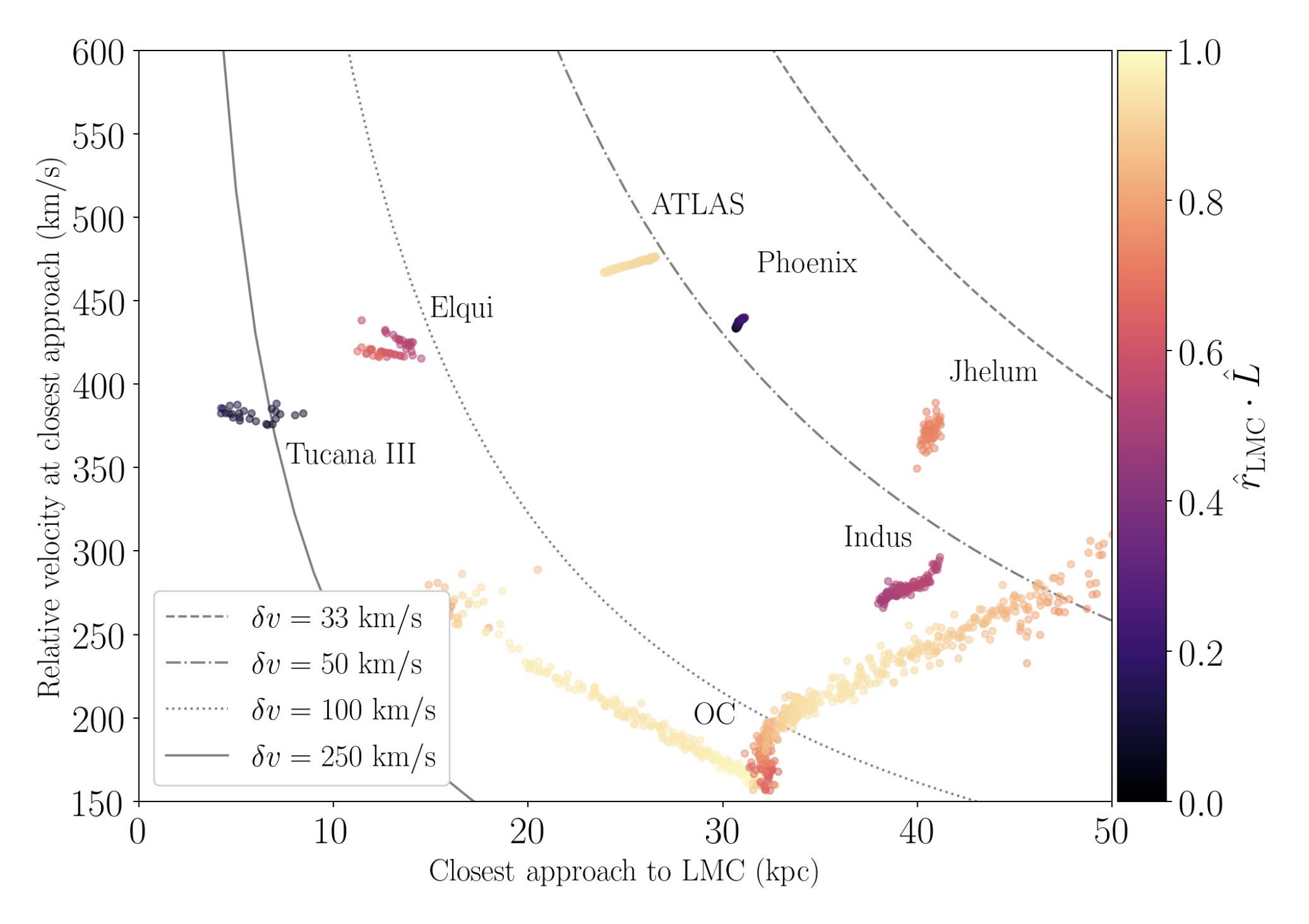}
\caption{Predicted perturbation by the LMC on each stream. The x-axis shows the distance of closest approach, and the y-axis is the relative velocity at closest approach. For each stream, we plot points spaced by $0.1 \deg$ in $\phi_1$. The curves represent lines of constant perturbation, assuming the interaction is impulsive. The color represents the component of the velocity kick in the direction of the angular momentum vector. Kicks in this direction present as offsets between the stream track and the proper motion direction, which are most easily measurable given currently available data.}
\label{fig:lmc_impact}
\end{figure*}

Given the orbits of each stream, we can predict the magnitude of the perturbation on the system by the LMC. Streams that pass close to the LMC with a small relative velocity are predicted to experience the strongest perturbations. For each stream, \tabref{results} lists the distance of closest approach to the LMC, the relative velocity between the stream and the LMC at closest approach, and the time at which this interaction occurs. The $95\%$ confidence interval on each LMC mass measurement is calculated from the highest density interval of the posterior distribution. With these parameters, we can predict the magnitude of the perturbation, as illustrated in Figure \ref{fig:lmc_impact}.

Figure \ref{fig:lmc_impact} shows the distance of closest approach and the relative velocity for several points along each stream, spaced by $0.1 \deg$ in $\phi_1$. The curves represent lines of constant perturbation strength, assuming the interaction is impulsive, using the results of Section 3.1 of \cite{Erkal:2015},
\begin{equation}
    \Delta v = \frac{G M}{b w},
\end{equation}
where $b$ is the impact parameter, $w$ is the relative velocity, and $\Delta v$ is the velocity kick to the stream. Streams that pass very close to the LMC with a small relative velocity are predicted to experience the largest perturbation, therefore Tuc III, Elqui, and OC should experience a more significant perturbation by the LMC than ATLAS, Phoenix, Indus, and Jhelum.

As described above, the interaction geometry is also an important consideration in predicting the effect of the LMC on each stream, since different perturbation geometries will affect different observables. Given the precise measurements of the proper motions, radial velocities, and, in particular, the stream tracks from \Gaia and \SSSSS, and the relative imprecision of the distance measurements, we will most easily be able to identify perturbations manifesting as changes to the stream proper motion and track.

\begin{figure}[htb!]
\centering
\includegraphics[width=\columnwidth]{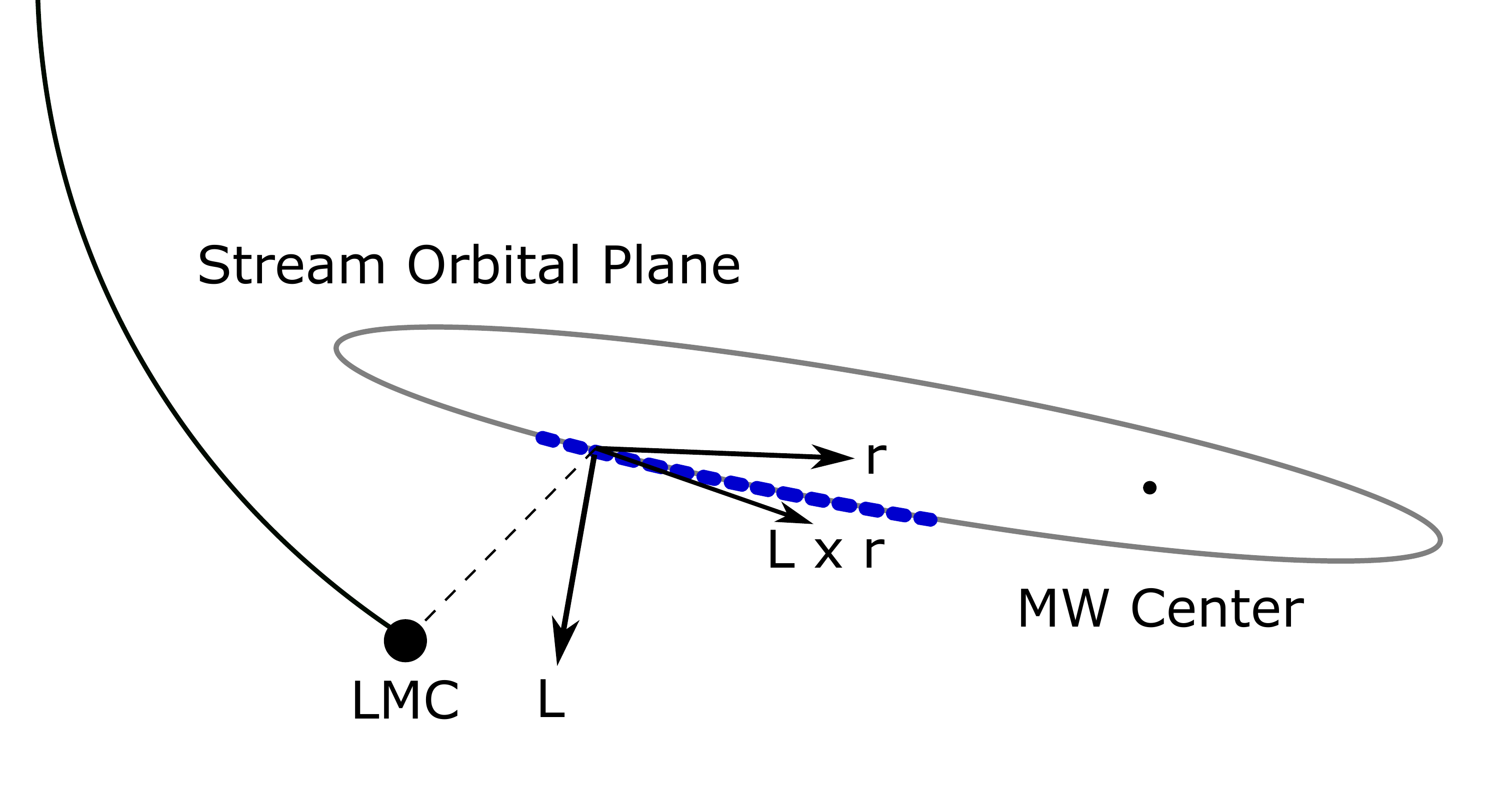}
\caption{The interaction geometry between a stream and the LMC. The blue dotted line represents a stream within an orbital plane traced by the solid black oval. The black dashed line represents the vector between the stream and the LMC at closest approach. We decompose this vector into components aligned with the angular momentum vector of the stream's orbit ($\hat{L}$), the radial vector between the stream and the Galactic center ($\hat{r}$), and a third perpendicular vector tangential to the stream's orbit ($\hat{L} \times \hat{r}$).}
\label{fig:lmc_diagram}
\end{figure}

We can characterize the observable effect of the LMC on a given stream by approximating the perturbation as a velocity kick towards the LMC at the point of closest approach. We then break down this velocity kick into components along the direction of the angular momentum, radial, and tangential vectors of the stream's orbit with respect to the Milky Way. This geometry is illustrated in Figure \ref{fig:lmc_diagram}. Velocity kicks out of the orbital plane, along the direction of the angular momentum vector of the stream's orbit, will produce an offset between the track of the stream and the direction of the proper motion, which can be precisely measured with the \SSSSS and \Gaia data. Kicks in the radial direction will manifest as an offset between the radial velocity and the distance gradient, which is more difficult to measure due to the difficulty in measuring distances along each stream. Perturbations aligned with the stream's velocity will be the most difficult to measure.

\begin{figure*}[htb!]
\centering
\includegraphics[width=0.8\textwidth]{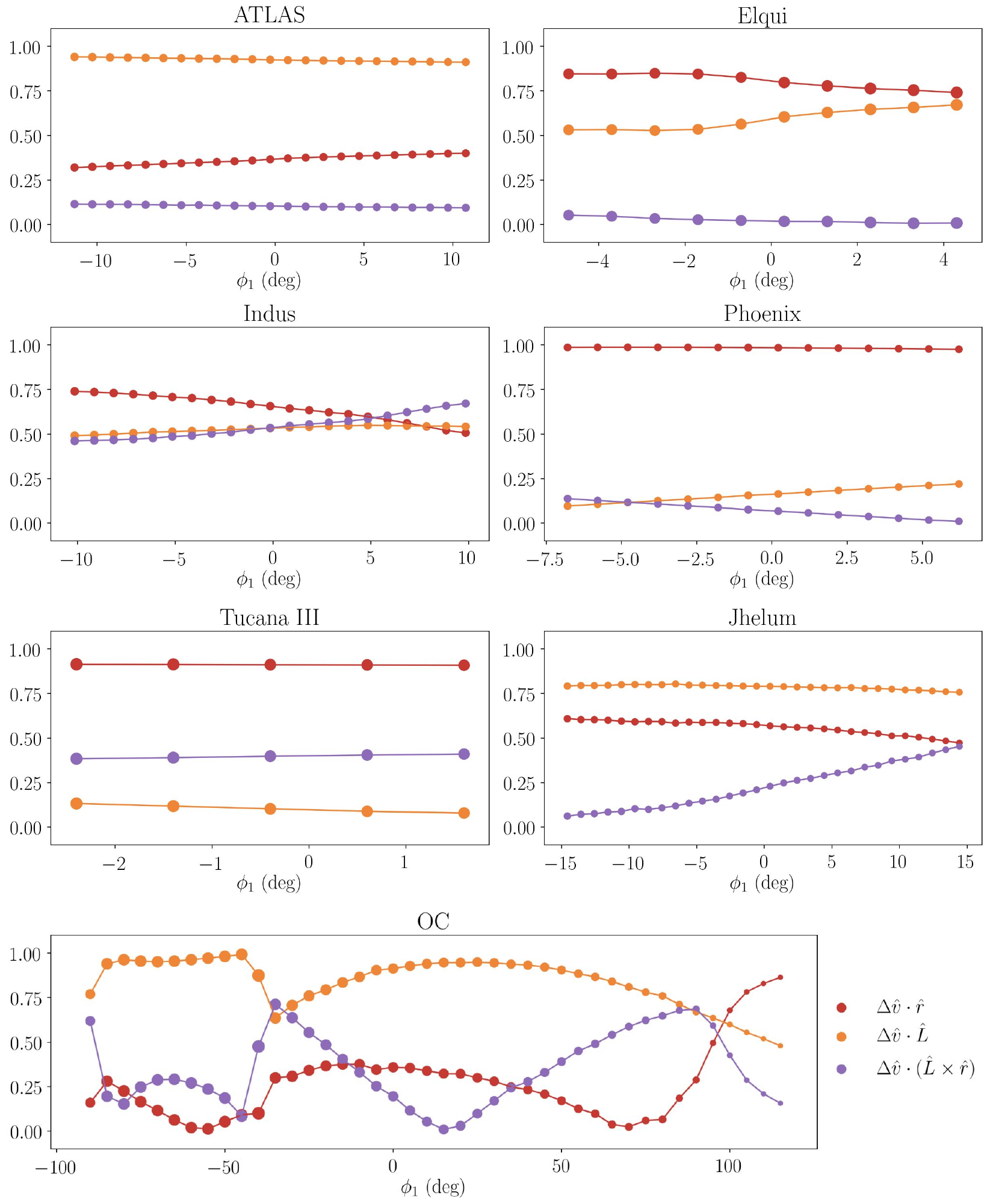}
\caption{Velocity kicks in three directions for each stream, assuming an impulsive interaction. Yellow represents the direction of the angular momentum of the stream orbit, red represents the radial direction (towards the center of the Milky Way) and purple represents the direction tangential to the stream orbit. Offsets in the angular momentum direction are the most visible given currently available data. Therefore, streams with large offsets in that direction, such as OC, are predicted to provide the strongest constraint on the LMC mass.}
\label{fig:kick_dirs}
\end{figure*}

The color scale in Figure \ref{fig:lmc_impact} represents the projection of the predicted perturbation in the angular momentum direction. Since perturbations in this direction are the most easily observable, this is a proxy for how observable the LMC's effect will be for the streams in our data set. Figure \ref{fig:kick_dirs} gives a more detailed view of the three components of the predicted perturbations.
Streams with the largest kicks in the angular momentum direction (yellow) are predicted to provide the strongest constraints on the LMC mass, while streams with small perturbations overall, or where the perturbations are primarily in the radial or tangential directions will provide weaker constraints. Interestingly, OC and ATLAS have the largest predicted perturbations out of the orbital plane, which coincides with the significant proper motion offsets measured for these streams \citep[e.g.,][]{Erkal:2019,Koposov:2019,Shipp:2019,Li:2021}. 

\subsection{Offsets} \label{sec:offsets}

\begin{figure*}[ht!]
    \centering
    \includegraphics[width=0.8\textwidth]{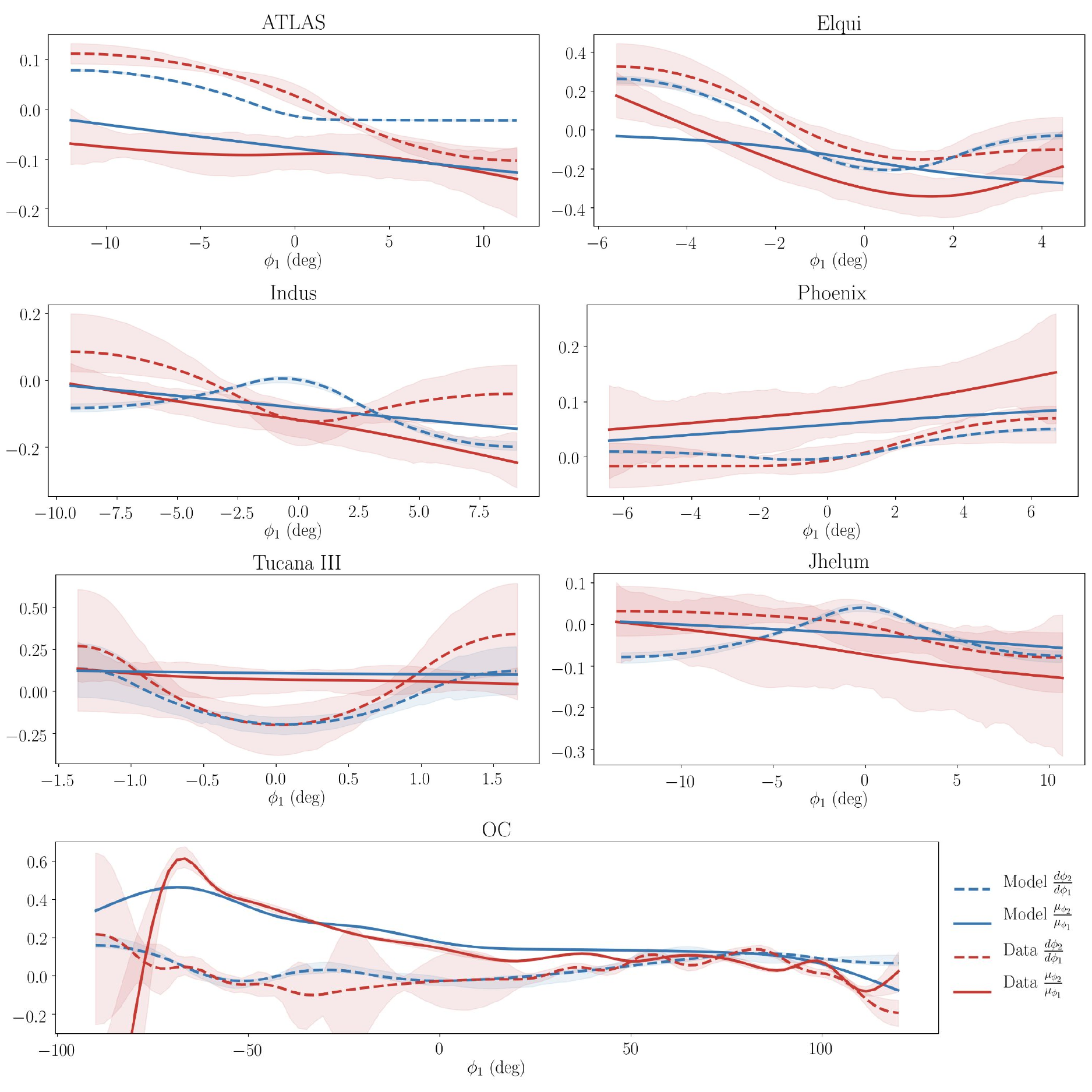}
    \caption{Offsets between the proper motion and the track of each stream. Solid lines represent the ratio of the proper motions ($\mu_{\rm \phi_2}/\mu_{\rm \phi_1}$), and the dashed lines represent the slope of the track on the sky. Blue lines/shaded regions correspond to the best-fit model, and red lines/shaded regions correspond to the \SSSSS data. The shaded regions represent the $1 \sigma$ uncertainties on each curve. For an unperturbed stream on a simple orbit, the dashed and solid lines will be aligned. However, several of these streams show some offset. OC, as predicted, has the largest proper motion offset, in both the data and the model.}
    \label{fig:pm_offsets} 
\end{figure*}

\begin{figure*}[ht!]
    \centering
    \includegraphics[width=0.8\textwidth]{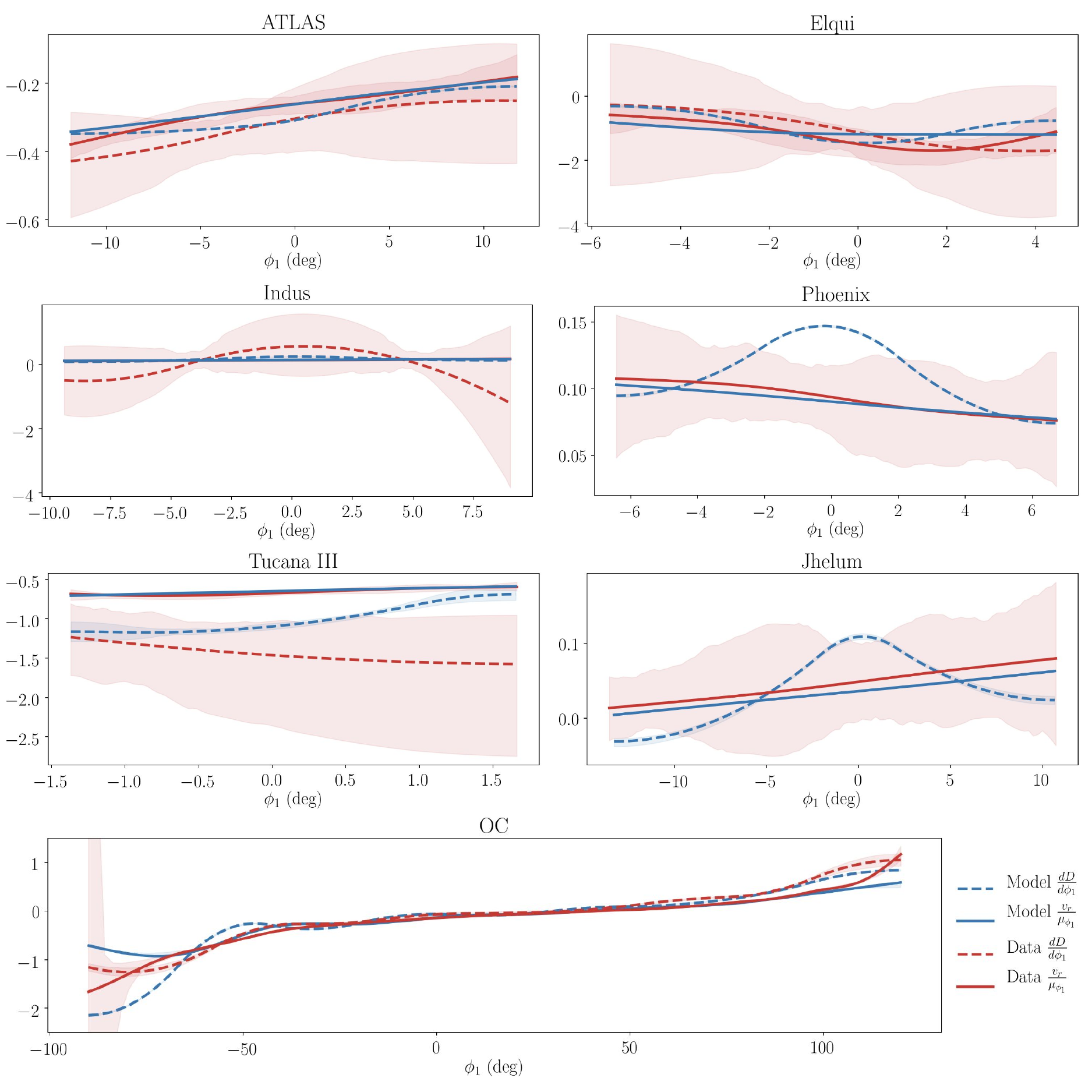}
    \caption{Offsets between the radial velocity and distance gradient of each stream. Solid lines represent the ratio of the solar reflex-corrected radial velocity to the proper motion ($v_r / \mu_{\rm \phi_1}$) along each stream, and the dashed lines represent the distance gradient. As in the above figure, blue lines correspond to the model, and red lines correspond to the data. We exclude the distance gradients for the Phoenix and Jhelum data due to large uncertainties. None of the streams have sufficient data to measure a significant offset between the two red lines. Tuc III is the only stream with a large predicted offset in the model. We find that the magnitude of this offset increases with larger LMC masses.}
    \label{fig:vr_offsets}
\end{figure*}

In order to examine the predicted and measured effect of the LMC on each stream we compare the orientation of the stream to the direction of motion \citep[e.g.,][]{Erkal:2019,deBoer2020,Li:2021}. Unperturbed streams roughly follow simple orbits and move in the direction in which they are extended. By combining \Gaia and \SSSSS data, we can compare the direction of motion to the extension of each stream in two ways. First, we can compare the track of the stream on the sky to the direction of its proper motion. Second, we can compare the distance gradient along the stream (along $\phi_1$) to the ratio of the radial velocity and the proper motion in the $\phi_1$ direction. These comparisons are illustrated in Figures \ref{fig:pm_offsets} and \ref{fig:vr_offsets}. The lines in these figures are each fit using cubic splines, following the MCMC method described in Section 3.1 of \citet{Erkal:2017}. The shaded bands represent the $1 \sigma$ uncertainties resulting from the MCMC spline fit.

Figure \ref{fig:pm_offsets} shows the offsets between the stream tracks and the direction of the proper motion for each stream in the model (blue) and the data (red). In each figure the dashed lines show the slope of the stream track ($\frac{d\phi_2}{d\phi_1}$), and the solid lines represent the ratio of the reflex-corrected proper motions ($\mu_{\rm \phi_{2}}/\mu_{\rm \phi_{1}}$) along the stream. We note that in this ratio we have used the quantity $\mu_{\rm \phi_1} = \frac{d\phi_1}{dt}$, which does not have a $\cos \phi_2$ term. Unperturbed streams should have no offset between the solid and dashed lines, while streams perturbed out of their orbital plane are predicted to have a significant offset. The offsets in the models roughly match the scale of the corresponding offsets in the data. Differences in the shapes of the curves may be due to lack of complexity in the model (e.g., additional small-scale perturbations), or the effect of placing the progenitor at $\phi_1 = 0\degr$. In the data (red), OC, ATLAS, and Elqui have the largest offsets between the two lines (note the differing axis scales between panels), which is consistent with the predictions illustrated in Figure \ref{fig:kick_dirs}. 

Figure \ref{fig:vr_offsets} shows the offsets in the radial direction. Here, the dashed line represents the distance gradient along the stream ($\frac{dr}{d\phi_1}$), while the solid line represents the ratio of the reflex-corrected radial velocity to the reflex-corrected proper motion along the stream ($v_r/\mu_{\rm \phi_1}$). Once again, unperturbed streams should show no offsets between these two lines, while streams perturbed in the radial direction should show significant offsets. The stream with the largest predicted offset in the model is Tuc III. This suggests that with improved distance measurements, and accounting for the possible effect of the Milky Way bar, we may be able to use Tuc III to place strong constraints on the mass of the LMC.

\subsection{Individual Streams}
\label{sec:individual}

\subsubsection{ATLAS}
\label{sec:atlas}
The ATLAS stream is a narrow stellar stream that was first discovered in the VST ATLAS survey \citep{Koposov:2014}, and further studied with deeper data from DES \citep{Shipp:2018}, and with spectroscopic data from \SSSSS \citep{Li:2021}. 

Our model of the ATLAS stream is fit to the \CHECK{72} \SSSSS members identified by \citet{Li:2021}. In addition, we include distance measurements to \CHECK{13} BHBs and \CHECK{5} RRLs in the calculation of the likelihood. The best-fit model for ATLAS is shown in Figure \ref{fig:atlas}. As noted in \secref{method}, we exclude Aliqa Uma, which has been shown to be an extension of the ATLAS stream \citep{Li:2021}, separated by a small perturber which has passed close to the stream.

Our fit to the ATLAS stream provides a measurement of $M_{\rm LMC} = 14.3^{+ 6.7}_{- 3.5}  \times 10^{10} \Msun$. \citet{Li:2021} also fit a model to the ATLAS stream, including an LMC with a fixed mass of $15 \times 10^{10} \Msun$. This is within $1 \sigma$ of our LMC mass measurement, and it is therefore unsurprising that \citet{Li:2021} obtained a good fit to the ATLAS stream including this fixed LMC mass and that their derived orbital properties are consistent with our best-fit orbit.

\figref{lmc_impact} and \figref{kick_dirs} show that ATLAS has one of the smaller predicted total perturbations, but one of the larger perturbations out of its orbital plane, suggesting that the majority of the perturbation should be observable as an offset between the track and proper motion of ATLAS. This is consistent with the large observed proper motion offset in ATLAS \citep{Shipp:2019, Li:2021}, and explains why ATLAS provides one of the strongest constraints on the LMC mass.

\subsubsection{OC}
\label{sec:chenab}
The northern part of the OC stream was discovered in SDSS by \citet{Grillmair:2006} and \citet{Belokurov:2006}. Chenab, the southern portion of OC, was then discovered in DES by \citet{Shipp:2018} and was later determined to be an extension of the Orphan stream by \citet{Koposov:2019}. Here we fit the full observed stream, using splines fit to data from \SSSSS, LAMOST, and APOGEE (Koposov et al. in prep.).

The southern component of OC has one of the largest predicted total perturbations by the LMC, and the largest along its angular momentum vector. Therefore, OC is expected to provide the strongest constraint on the LMC mass of the 5 streams included in this work. As seen in \figref{lmc_mass}, OC does in fact provide one of the tightest constraints.

\figref{lmc_impact} illustrates how the impact of the LMC on OC varies along the length of the stream. The strongest predicted impact is along the southern portion of the stream. This part of the stream is also most significantly perturbed out of its orbital plane (\figref{kick_dirs}), which is consistent with the large proper motion offset seen in \figref{pm_offsets}. \citet{Fardal:2019} identify a misalignment between the spatial track and velocity vector along the northernmost component of the stream, which is consistent with the offset shown at large $\phi_1$ in \figref{vr_offsets}. At $\phi_1 > 125 \degr$, we see an offset in both the data and the model between the distance gradient and the ratio of the radial velocity to the proper motion along the stream, which suggests the stream is curving away in distance at a greater rate than would be expected along an unperturbed orbit.

\CHECK{We obtain a best-fit value of \CHECK{$M_{\rm LMC} = 18.8^{+ 3.5}_{- 4.0} \times 10^{10} \Msun$}, which differs from the result of \citet{Erkal:2019} by $\roughly 1 \sigma$. This offset may be due to differences in dataset or model. We discuss the possible sources of this difference in greater detail in \secref{disc}.}

\CHECK{In fitting the OC stream, we find the resulting LMC mass depends somewhat on the placement of the progenitor along the stream. Placing the progenitor along the portion of the stream discovered in the DES footprint resulted in a larger LMC mass. However, the total enclosed LMC mass within the closest passage of the progenitor in each case changes only slightly. This is due in part to the fact that we fix the scale radius of the LMC in order to match the circular velocity measurement from \citet{vanderMarel:2014}. The influence of the progenitor placement may therefore be mitigated by introducing more flexibility in the outskirts of the LMC potential.}

\subsubsection{Elqui}
\label{sec:elqui}
Elqui was discovered by \citet{Shipp:2018} in the DES data, and is the most distant of the DES stellar streams at 50 \kpc. Our list of likely Elqui members includes \CHECK{43} \SSSSS stars, with \CHECK{3} BHBs and \CHECK{5} RRLs, which we fit to constrain the mass of the LMC to $M_{\rm LMC} = 16.8^{+ 5.2}_{- 3.0} \times 10^{10} \Msun$. Elqui passes within 15\kpc of the LMC and shows signs of significant perturbation. In particular, the track of Elqui appears to deviate from the great circle orbit that passes through the two endpoints (Figure \ref{fig:models}). This deviation can be reproduced by a model including a massive LMC, but cannot be reproduced if the LMC is excluded. Elqui has a significant predicted perturbation in the angular momentum direction, but the largest predicted velocity kick is in the radial direction. This suggests that with improved distance measurements we may be able to use Elqui to place an even stronger constraint on the LMC mass.

\CHECK{Elqui is measured to have a large velocity dispersion of $16.8 \kms$ (Li et al., in prep). For comparison, streams with similar predicted progenitor masses, such as Indus, have typical velocity dispersions of $< 10 \kms$.} This large dispersion is also present in the best-fit stream model. This is despite the fact that our progenitor for Elqui is dynamically cold, with a mass of only $M_{\rm prog} = 10^6 \Msun$. This indicates that the measured stream dispersion does not necessarily map directly onto the dispersion of the progenitor system. It seems unlikely that this unusually large dispersion is due to perturbation by the LMC, as it does not significantly decrease with decreasing LMC mass. The large dispersion may be due to the high orbital eccentricity of Elqui. The source of this large dispersion requires further investigation.

\citet{Shipp:2018} proposed a possible connection between Elqui and the Magellanic stream, due to the similar distance and orientation of Elqui. \citet{Shipp:2019} showed that the proper motion of Elqui is inconsistent with an association with the Magellanic stream. We further support this conclusion here, noting that at closest approach, Elqui has a velocity relative to the LMC of $\sim 420 \kms$, indicating that it cannot be bound to the Magellanic system.

\subsubsection{Indus}
\label{sec:indus}
The Indus stream is a thick stream discovered by \citet{Shipp:2018} in DES. Our list of likely Indus members consists of \CHECK{59} \SSSSS stars, including \CHECK{4} BHBs and \CHECK{1} RRL, which we fit to measure an LMC mass of \CHECK{$M_{\rm LMC} = 15.6^{+ 8.6}_{- 3.6} \times 10^{10} \Msun$}. Of the 5 streams used to fit the LMC mass, Indus has the largest distance of closest approach to the LMC. However, it passes by at a slow relative velocity and with a geometry such that the predicted velocity kick in the angular momentum direction is similar in magnitude to that of ATLAS or Elqui. Indus is predicted to have experienced a radial perturbation equal in magnitude to its perturbation out of its orbital plane. Indus also has only a small number of known distance tracers, and therefore a large uncertainty in its distance gradient. Given the large width of Indus, it is likely that additional distance tracers lie outside the limits of the \SSSSS footprint. With an improved measurement of the distance gradient, it is likely that Indus can provide an even stronger constraint on the LMC mass.

\subsubsection{Phoenix}
\label{sec:phoenix}
Phoenix is a narrow stream discovered in the DES data by \citet{Balbinot:2016} and further studied by \citet{Shipp:2018}. Significant density variations of unknown origin have been identified along the stream \citep[][Tavangar et al. in prep.]{Balbinot:2016}. These variations are unlikely to be due to a large-scale perturbation by a system like the LMC. \citet{Wan:2020} found that Phoenix has a very low metallicity ($\feh \approx -2.7$\,dex), suggesting that Phoenix represents the tidal debris of the most metal-poor Milky Way globular cluster known to date.

Phoenix has the least sensitivity to the LMC mass of the five streams, as seen in Figure \ref{fig:lmc_mass}. We fit the \CHECK{30} \SSSSS members of Phoenix, as published in \citet{Wan:2020}, and \CHECK{3} BHBs to obtain an LMC mass measurement of \CHECK{$M_{\rm LMC} = 2.7^{+ 8.5}_{- 0.7} \times 10^{10} \Msun$}.
It is unsurprising that Phoenix provides a weaker constraint on the LMC mass than the other streams, for two primary reasons. First, the Phoenix stream has the smallest predicted velocity kick from the LMC, as shown in Figure \ref{fig:lmc_impact}. Only Indus has a larger distance of closest approach to the LMC, and Phoenix passes by the LMC with a much larger relative velocity. Second, the orientation of the Phoenix stream relative to the LMC suggests that the majority of the velocity kick is in the radial direction, as seen in Figure \ref{fig:kick_dirs}, and is thus very difficult to detect with current observations. This suggests that with an improved distance gradient we may be better able to measure the effect of the LMC on the Phoenix stream, and perhaps tighten the constraint on the mass of the LMC.

\subsubsection{Tucana III}
\label{sec:tuciii}
Tuc III was discovered in the second year of DES data by \citet{Drlica-Wagner:2015} and consists of thin tidal tails extending from a central progenitor. We exclude Tuc III from this analysis due to its probable interaction with the Milky Way bar, as discussed in \secref{method}. However, Tuc III is an interesting stream to consider in the context of the LMC since it passes within $10 \kpc$ of the LMC. We fit a model to Tuc III including an LMC with a mass fixed to $15 \times 10^{10} \Msun$, motivated by the results of \citet{Erkal:2019}. 

\citet{Erkal:2018a} modeled the Tuc III stream and predicted the proper motion offset that would be revealed with the release of \Gaia DR2. They predicted a large value of the reflex corrected proper motion perpendicular to the stream, $\mu_{\rm \phi_2}$, would be observed for any LMC mass greater than $\roughly 10^{10} \Msun$. However, the observed proper motion is found to be generally aligned with the track of the stream \citep{Shipp:2019}. This inconsistency may be a result of how the Milky Way potential was modeled by \citet{Erkal:2018a}. The LMC is known to induce a significant reflex motion in the Milky Way \citep[e.g.,][]{Gomez:2015,Garavito-Camargo:2019,Erkal2021,Petersen:2021}; however, \citet{Erkal:2018a} fixed the centroid of the Milky Way potential, which may have biased the fits to Tuc III and thereby the predictions for $\mu_{\phi_2}$. In this paper, we allow the center of the Milky Way to move in response to the LMC, and find that we are able to obtain a good fit to the data without a significant proper motion offset. Instead, we find that Tuc III is predicted to have a large offset between its radial velocity and distance gradient (see \figref{vr_offsets}). This predicted offset suggests that a precise measurement of the distance gradient along Tuc III, along with an understanding of the effects of the Milky Way bar, may enable strong constraints on the LMC mass. However, improved distance gradient measurements are difficult due to the distance and low luminosity of the stream. Therefore, despite Tuc III's close passage to the LMC its utility as a probe of the LMC mass is unclear.

\subsubsection{Jhelum}
\label{sec:jhelum}
Jhelum is a thick stream discovered in the DES data by \citet{Shipp:2018}. As with Tuc III, we have excluded Jhelum from our analysis due to the evidence of perturbation presented by \citet{Bonaca:2019} and \citet{Shipp:2019}. In particular, \citet{Bonaca:2019} identified two distinct spatial components in Jhelum---a thin dense component and a broader, more diffuse component---while \citet{Shipp:2019} identified two distinct proper motion components in Jhelum in the \Gaia data. The cause of this unique morphology remains unknown. Despite this evidence of perturbation, we fit a model to Jhelum with a fixed LMC mass \CHECK{of $15 \times 10^{10} \Msun$} in order to explore the predicted impact of the LMC. We find that Jhelum is not predicted to be significantly perturbed by the LMC.

\subsection{Combined Measurement}
\label{sec:combined}
This work is the first step towards using a population of streams to measure the potential of the LMC. Ultimately, we would like to combine the measurements of multiple stellar streams in order to increase the precision of our measurement of the LMC potential.   However, there are several challenges associated with combining the measurements presented here. First, we chose to fit the streams individually in order to examine the effect of the LMC on each individual system and to develop an understanding of how different streams are perturbed by the LMC depending on their relative orbits and times of closest approach. In addition, combining posterior distributions leads to overcounting of the priors that were repeated in each fit. With these caveats in mind, and primarily for illustrative purposes, we can combine the individual stream measurements by taking the product of Kernel Density Estimates (KDEs) fit to the posterior distributions of the individual streams. From this simple procedure, we derive a combined measurement of the LMC mass of $18.4^{+1.8}_{-1.9} \times 10^{10} \Msun$.  This measurement differs from the result of \citet{Erkal:2019} by $1.3 \sigma$ and the result of \citet{Vasiliev:2021} by $1.5 \sigma$. \CHECK{In addition, by combining measurements from multiple streams, we obtain uncertainties on the LMC mass that are smaller than those of either of the previous measurements.}
In the future, we will simultaneously model a population of streams in order to fit an aspherical, deforming LMC potential, e.g., with basis function expansions \citep{Garavito-Camargo:2020}.

\begin{figure*}
\centering
\includegraphics[width=0.9\textwidth]{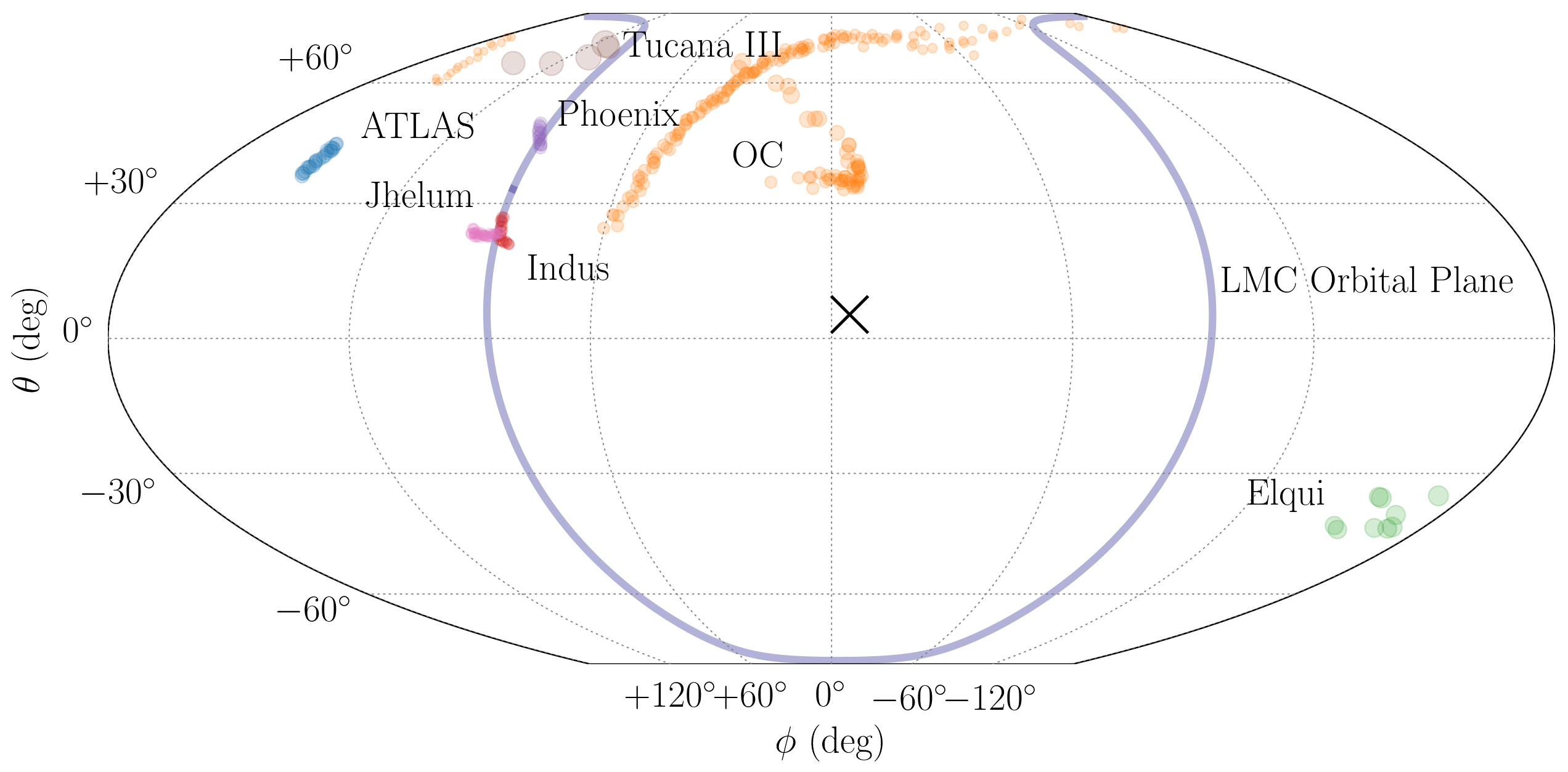}
\caption{The position around the LMC at which each stream passes at closest approach. $\theta$ and $\phi$ represent the polar angles of the closest approach of each stream as viewed from the center of the LMC. The coordinate system is identical to Galactocentric coordinates shifted to the center of the LMC. The cross shows the direction of the LMC's orbital angular momentum and the blue line shows the plane of the LMC's orbit around the Milky Way. The points are spaced by $0.1 \deg$ in $\phi_1$ along the stream, and the size of the points is inversely proportional to the distance of closest approach to the LMC. Most of the streams pass to the north of the LMC, while Elqui passes below the galaxy. Each stream probes a different part of the LMC potential, by passing by at a different position, distance, and time. Ultimately, we will be able to use a large population of streams to constrain the disrupting, asymmetrical potential of the LMC.}
\label{fig:lmc_positions}
\end{figure*}

\section{Discussion}
\label{sec:disc}

We have fit the mass of the LMC independently using 5 stellar streams with data from \SSSSS, \Gaia EDR3, and DES DR1.
In \secref{results}, we present the constraint on the LMC mass from each of these fits (Figure \ref{fig:lmc_mass}), and discuss the unique interactions between each stream and the LMC (Figures \ref{fig:lmc_impact} and \ref{fig:kick_dirs}). Here, we place these measurements in context. We discuss how these measurements begin to establish a consistent picture of the influence of the Milky Way's largest satellite on the population of southern stellar streams, draw comparisons to other measurements, and discuss future efforts to model a realistic LMC and Milky Way potential.

As discussed in \secref{results}, due to the precision of the \Gaia EDR3 proper motion measurements, our data are most sensitive to perturbations out of the stream orbital plane, which produce observed misalignments between stream tracks and proper motions. Therefore, by combining the results of \figref{lmc_impact} and \figref{kick_dirs}, which demonstrate the predicted magnitude and geometry of the perturbation of the LMC on each stream, we can predict the ability of each stream to constrain the mass of the LMC. These predictions suggest that OC, followed by Elqui, ATLAS, and Indus should have the largest observed proper motion offsets and should provide the strongest constraints on the LMC mass. In fact, these are the streams that we find provide the strongest constraints. Streams with large offsets in other directions may require additional data in order to better constrain the LMC mass. For example, Tuc III, Elqui, and Phoenix all have large predicted offsets in the radial direction, which are detectable only with precise measurements of the distance gradient along the stream. The constraints on the LMC mass should show significant improvement with improved distance measurements, which may be obtained in the future with additional distance tracers, deeper photometric data, or the extension of precise parallax measurements to more distant sources. Given these predictions, it is also unsurprising that Phoenix does little to  constrain the LMC mass, given it has both the smallest predicted total perturbation, and is predicted to have been perturbed primarily in the radial direction.

By examining the orbits of each stream relative to the LMC, we have assembled a consistent picture of the effect of the LMC on this population of stellar streams. This consistency supports our measurements of a massive LMC with $M_{\rm LMC} \sim 1.8 \times 10^{11} \Msun$. We find that all of the streams considered in this work, with the exception of Phoenix, prefer a total mass of the LMC consistent with this value within $1 \sigma$, supporting an LMC with a mass of at least $15\%$ that of the Milky Way.

\CHECK{This measurement supports, and in some cases exacerbates, studies of the unusual nature of the Milky Way-LMC system. \citet{Dooley:2017a} found a discrepancy between the predicted and observed population of LMC satellites. Our measurement of $M_{\rm LMC} \sim 1.8 \times 10^{11} \Msun$ supports their predictions and further emphasizes the need for the continued study of the stellar mass function of satellites in the vicinity of the LMC. In addition, this measurement supports studies of the rarity of LMC-mass satellites around Milky Way-like hosts \citep{Busha:2011}, as well as predictions of the post-merger properties of the Milky Way and a massive LMC \citep{Cautun2019}. Furthermore, the LMC's impact on our sample of streams motivates the continued search for observational signs of its influence on other Milky Way structures \citep[e.g.][]{Belokurov2019,Conroy2021,Erkal2021,Petersen:2021,Ji:2021}. }

\subsection{Comparison to Previous Measurements}
\label{sec:comparison}

Prior to this work, two streams had been used to fit the mass of the LMC --- OC \citep{Erkal:2019} and Sagittarius \citep{Vasiliev:2021}. \citet{Erkal:2019} measured an LMC mass of $1.38^{+0.27}_{-0.24} \times 10^{11} \Msun$, and \citet{Vasiliev:2021} obtained a consistent measurement of $M_{\rm LMC} = 1.3 \pm 0.3 \times 10^{11} \Msun$.

\CHECK{As mentioned in \secref{individual}, our fit to OC is a particularly interesting comparison to the results of fitting the same stream by \citet{Erkal:2019}.}
In this work, OC prefers an LMC mass of \CHECK{$18.8^{+ 4.0}_{- 3.5} \times 10^{10} \Msun $}, which differs from the result of the \citet{Erkal:2019} by \CHECK{$\roughly 1 \sigma$}. 
\CHECK{This may be due in part to the fact that we are fitting a different dataset, and in particular that \citet{Erkal:2019} did not include radial velocities in their likelihood. In fact, the \citet{Erkal:2019} best-fit model does not match the \SSSSS radial velocity measurements along the southern portion of the stream.}

Another important distinction is that we do not vary the Milky Way potential, whereas \citet{Erkal:2019} allow the mass, shape, and orientation of the Milky Way to vary; they find that the OC stream prefers a substantially asymmetric halo shape. Similarly, \citet{Vasiliev:2021} find that the Sagittarius stream prefers a twisted dark matter halo that is aligned with the Milky Way disk in the inner halo but flattened in an almost perpendicular direction in the outskirts. This suggests that further work on simultaneously fitting multiple streams in a flexible joint Milky Way and LMC potential is necessary in order to entangle the complex interplay between the LMC and Milky Way potentials.

\subsection{Shape of the LMC} 
\label{sec:shape}

The LMC is also known to have an asymmetric potential and in fact is disrupting as it orbits the Milky Way \citep[e.g.,][]{Garavito-Camargo:2019,Garavito-Camargo:2020,Petersen:2020}. In this work, we have used a simplified model of the LMC. We have modeled the potential of the LMC as a spherical Hernquist profile \citep{Hernquist:1990}, while in reality the LMC potential is more complex. A more realistic model of the LMC potential is necessary to explore the full complexity of the interactions between the LMC and Milky Way stellar streams. The fact that the LMC is a complex, non-spherical, disrupting system suggests that the \SSSSS streams may experience very different effects from the LMC depending on which part of the galaxy they pass by and at what time. 

To explore this idea further, Figure \ref{fig:lmc_positions} shows the predicted point of closest approach of each stream relative to the LMC. Each stream probes a different part of the LMC potential, and notably Elqui passes on the opposite side of the LMC from the other streams. In this figure we also show the orbital plane of the LMC around the Milky Way (blue line). Since the LMC material spreads out the most within this plane \citep[e.g. see Fig. 10 in][]{Erkal:2019}, the streams closest to this plane likely experience a stronger LMC force field than those farther away. As we continue to build up our dataset, our population of streams, and the complexity of our models, we should be able to use each of these streams to measure not only the total mass of the LMC, but its disrupting, asymmetrical radial profile. 

\section{Conclusions}
\label{sec:conc}

The dynamical influence of the LMC on structures in the Milky Way has been the focus of much recent work \citep[e.g.,][]{Erkal:2018a,Erkal:2019,Erkal2021,Erkal2020_sats,Vasiliev:2021,Petersen:2020}. Here, we measure the LMC mass with five streams in the Southern Galactic hemisphere  with full 6D phase space measurements. These streams are sensitive to the LMC due to their proximity. Previous work  has shown that many of these streams have proper motions misaligned with their stream tracks \citep{Shipp:2019}, which is a tell-tale sign of a gravitational perturbation \citep[e.g.,][]{Erkal:2019}. We find the mass of the LMC to be $\sim1.8\times10^{11} \Msun$, consistent with previous measurements of the LMC mass with stellar streams \citep{Erkal:2019, Vasiliev:2021}. By examining the interaction between each individual stream and the LMC, we can build a consistent picture of how the Milky Way's largest satellite has perturbed the population of stellar streams.

In order to understand the constraining power of each stream, we used a simple model which assumes the velocity kick imparted by the LMC is impulsive. We then decomposed the predicted velocity kicks into three directions relative to each stream's orbital plane: aligned with the angular momentum direction (i.e., perpendicular to the stream plane), in the radial direction (i.e., towards the Milky Way), and the tangential direction (i.e., along the stream). We found that the streams with the most stringent constraints on the LMC mass have significant velocity kicks which are perpendicular to the stream plane since these are visible as a misalignment between the proper motion direction and stream track. 

Our best-fit models predict that several streams (Tuc III, Elqui, and Phoenix) are expected to have a significant misalignment along the line of sight. This alignment should be observable by comparing their distance gradient with the ratio of the radial velocity to the proper motion along the stream. These are currently not measurable due to the relatively large uncertainty in distance compared to the other observables. With improved measurements of radial velocities and distance gradients in the future we should be able to more precisely measure the impact of the LMC in the radial direction. The detection of this misalignment will allow us to better constrain the LMC mass with these streams. 

Although the main result of this work is that we can successfully model all of the streams with an LMC of mass $\roughly 1.8\times10^{11} \Msun$, there are a number of additional avenues for exploration. First, we have varied the LMC mass but we have kept the Milky Way fixed. Fitting these streams, along with streams from the Northern Galactic hemisphere, will likely provide strong constraints on the shape of the Milky Way's dark matter halo. Second, we have neglected the tidal deformation of the Milky Way and LMC which may have a substantial effect on these streams. Interestingly, the streams considered in this work approach the LMC from a variety of directions (see Fig. \ref{fig:lmc_positions}), suggesting that they will be a powerful probe of any deformations of the LMC. 

Furthermore, in this work we fit the mass of the LMC with 5 of the stellar streams observed in the first year of \SSSSS observations. \SSSSS has now measured radial velocities of more than 20 stellar streams and observations are ongoing. In addition, future photometric surveys such as the Vera C. Rubin Observatory Legacy Survey of Space and Time (LSST) will observe more distant streams, many of which will have passed on the opposite side of the LMC (see Elqui in \figref{lmc_positions}), and will provide a more complete view of the potential of the LMC.

Further study of the complexity of the Milky Way and LMC potential with a large population of stellar streams will build upon this work to reveal a more complete picture of the effect of the Milky Way's largest satellite on our Galaxy.

\section{Acknowledgments}

We thank Paul McMillan for providing the MCMC chains from his analysis in \cite{McMillan:2017}. This paper is based upon work that is supported by the Visiting Scholars Award Program of the Fermilab Universities Research Association. NS thanks the LSSTC Data Science Fellowship Program, her time as a Fellow has benefited this work.
TSL is supported by NASA through Hubble Fellowship grant HF2-51439.001 awarded by the Space Telescope Science Institute, which is operated by the Association of Universities for Research in Astronomy, Inc., for NASA, under contract NAS5-26555.
ABP is supported by NSF grant AST-1813881. This research has been supported in part by the Australian Research Council Centre of Excellence for All Sky Astrophysics in 3 Dimensions (ASTRO 3D) through project number CE170100013.
This paper includes data obtained with the Anglo-Australian Telescope in Australia. We acknowledge the traditional owners of the land on which the AAT stands, the Gamilaraay people, and pay our respects to elders past and present.
This work has made use of data from the European Space Agency (ESA) mission
{\it Gaia} (\url{https://www.cosmos.esa.int/gaia}), processed by the {\it Gaia}
Data Processing and Analysis Consortium (DPAC,
\url{https://www.cosmos.esa.int/web/gaia/dpac/consortium}). Funding for the DPAC
has been provided by national institutions, in particular the institutions
participating in the {\it Gaia} Multilateral Agreement.

\bibliographystyle{aasjournal}
\bibliography{main}

\appendix
\numberwithin{figure}{section}
\numberwithin{table}{section}

\section{Model Parameters}
\label{app:params}

\newcommand{\paramscaption}{Stream progenitor and LMC parameters.}
\newcommand{\paramscomments}{Progenitor and LMC parameters for the best-fit stream models. Note that for Tucana III and Jhelum, all LMC parameters were fixed. For all streams, $M_{\rm prog}$ and $r_{\rm s, prog}$ were fixed. See parameter descriptions in \tabref{priors}.}
\begin{deluxetable*}{l cccccccc}[htp]
\tablecolumns{13}
\tablewidth{0pt}
\tabletypesize{\scriptsize}
\tablecaption{ \paramscaption }
\label{tab:params}
\tablehead{ \colhead{Parameter} & \colhead{ATLAS} & \colhead{OC} & \colhead{Elqui} & \colhead{Indus} & \colhead{Phoenix} & \colhead{Tucana III} & \colhead{Jhelum} }
\startdata
$\phi_{\rm 2, prog}\ (\deg)$                      & 0.73 & \CHECK{-0.77} & 0.31 & 0.30 & -0.11 & -0.10 & 0.01 \\
$\sigma_{\rm \phi_2, prog}\ (\deg)$               & 0.30 & \CHECK{--} & 0.18 & 0.48 & 0.11 & 0.07 & 0.29 \\
$v_{\rm r, prog}\ (\kms)$                         & -110.09 & \CHECK{96.97} & -57.60 & -52.83 & 47.61 & -102.26 & -5.35 \\
$\sigma_{\rm v_r, prog}\ (\kms)$                  & 4.06 & \CHECK{--} & 13.60 & 4.83 & 3.14 & 0.89 & 18.41 \\
$d_{\rm prog}\ (\kpc)$                             & 21.22 & \CHECK{18.78} & 52.85 & 15.77 & 17.29 & 23.97 & 12.79 \\
$\mu_{\rm \phi_1 \star, prog}\ (\mathrm{mas\ yr^{-1}})$ & -0.40 & \CHECK{4.25} & -0.56 & -5.74 & -0.92 & 0.21 & -7.13 \\
$\mu_{\rm \phi_2, prog}\ (\mathrm{mas\ yr^{-1}})$ & -0.90 & \CHECK{1.95} & -0.32 & -1.27 & -2.55 & -1.60 & -3.15 \\
\tableline
$M_{\rm LMC}\ (10^{10} \Msun)$                    & 16.04 & \CHECK{18.18} & 17.95 & 17.21 & 6.93 & 15.00 & 15.00 \\
$v_{\rm r, LMC}\ (\kms)$                          & 262.51 & \CHECK{263.91} & 263.93 & 261.78 & 262.13 & 262.20 & 262.20 \\
$d_{\rm LMC}\ (\kpc)$                              & 49.48 & \CHECK{51.36} & 48.44 & 50.19 & 50.05 & 49.97 & 49.97 \\
$\mu_{\rm \alpha \star, LMC}\ (\mathrm{mas\ yr^{-1}})$  & 1.91 & \CHECK{1.92} & 1.91 & 1.91 & 1.91 & 1.91 & 1.91 \\
$\mu_{\rm \delta, LMC}\ (\mathrm{mas\ yr^{-1}})$  &  0.21 & \CHECK{0.36} & 0.17 & 0.22 & 0.23 & 0.23 & 0.23 \\
\tableline
$M_{\rm prog}\ (\Msun)$                   & $2 \times 10^4$ & $1 \times 10^7$ & $1 \times 10^6$ & $1 \times 10^7$ & $2 \times 10^4$ & $2 \times 10^3$ & $2 \times 10^7$ \\
$r_{\rm s, prog}\ (\kpc)$                         & 0.01 & 0.5 & 0.1 & 0.1 & 0.01 & 0.05 & 0.1  \\
\enddata
{\footnotesize \tablecomments{ \paramscomments }}
\end{deluxetable*}

\newcommand{\orbitcaption}{Stream orbital parameters.}
\newcommand{\orbitcomments}{
Orbital parameters of the best-fit stream models. $\phi$ and $\psi$ are the Galactocentric azimuthal and polar angles of the orbital pole, respectively.}
\begin{deluxetable*}{l cccccccc}[htp]
\tablecolumns{13}
\tablewidth{0pt}
\tabletypesize{\scriptsize}
\tablecaption{ \orbitcaption }
\label{tab:orbits}
\tablehead{ \colhead{Parameter} & \colhead{ATLAS} & \colhead{OC} & \colhead{Elqui} & \colhead{Indus} & \colhead{Phoenix} & \colhead{Tucana III} & \colhead{Jhelum} }
\startdata
$r_{\rm peri}\ (\kpc)$  & $12.7^{+0.2}_{-0.2}$  & $16.3^{+ 7.6}_{- 0.1}$  & $9.7^{+ 0.7}_{- 0.9}$   & $12.9^{+ 0.2}_{- 0.1}$  & $13.1^{+ 0.2}_{- 0.2}$  & $2.8^{+ 0.1}_{- 0.1}$  & $9.0^{+ 0.1}_{- 0.2}$  \\
$r_{\rm apo}\ (\kpc)$   & $40.0^{+0.4}_{-0.4}$  & $69.1^{+0.7}_{-0.8}$  & $66.9^{+ 3.2}_{- 2.2}$  & $19.9^{+ 1.8}_{- 1.0}$  & $18.5^{+ 0.1}_{- 0.1}$  & $44.3^{+ 5.1}_{- 0.7}$  & $29.5^{+ 0.5}_{- 1.8}$  \\
$\phi\ (\deg)$          & 349.2 & 190.4 & 344.7 & 125.7 & 61.8  & 322.9 & 119.5 \\
$\psi\ (\deg)$          & 114.9 & 137.8 & 90.9  & 110.4 & 119.9 & 71.5  & 97.7  \\
\enddata
{\footnotesize \tablecomments{ \orbitcomments }}
\end{deluxetable*}


\newcommand{\potentialcaption}{Potential parameters}
\newcommand{\potentialcomments}{Parameters of the potential from \cite{McMillan:2017} which we use to fit streams in this work. We note that the parameters here come from a realization of the posterior MCMC chains in \cite{McMillan:2017} and are thus consistent with that work. For ease of use, the parameters are in the same format as \cite{McMillan:2017}. }
\begin{deluxetable*}{l ccc}[htp]
\tablecolumns{13}
\tablewidth{0pt}
\tabletypesize{\scriptsize}
\tablecaption{ \potentialcaption }
\label{tab:potential}
\tablehead{\colhead{Parameter} &  &\colhead{Property} & 
}
\startdata
$\Sigma_{0, \rm thin}$ & 679.3 $\Msun$pc$^{-2}$ & $v_0$ & 233.7 km s$^{-1}$ \\
$R_{d, {\rm thin}}$ & $2.823$ kpc & $M_b$ & $9.821\times10^{9} \Msun$ \\
$\Sigma_{0, \rm thick}$ & 231.8 $\Msun$pc$^{-2}$ & $M_{d, {\rm thin}}$ & $3.403\times10^{10} \Msun$ \\
$R_{d, {\rm thick}}$ &  $2.956$ kpc & $M_{d, {\rm thick}}$ &  $1.272\times10^{10} \Msun$ \\
$\rho_{0,b}$ & 104.7 $\Msun$pc$^{-3}$ & $M_{\rm v}$ & $8.273\times10^{11} \Msun$ \\
$\rho_{0,h}$ & 0.01576 $\Msun$pc$^{-3}$ & $c_{v'}$ & 15.07  \\
$r_h$ & 13.14 kpc & &  \\
$R_0$ & 8.228 kpc & & \\
$U$ & 8.406 km s$^{-1}$ & & \\
$V$ & 12.01 km s$^{-1}$ & & \\
$W$ & 7.280 km s$^{-1}$ & & \\
\enddata
{\footnotesize \tablecomments{ \potentialcomments }}
\end{deluxetable*}

\FloatBarrier
\onecolumngrid 
\section{Stream Models} 
\label{app:models}

\begin{figure}[htp]
    \centering
    \gridline{\fig{ATLAS_29_42}{0.45\textwidth}{(a)}
              \fig{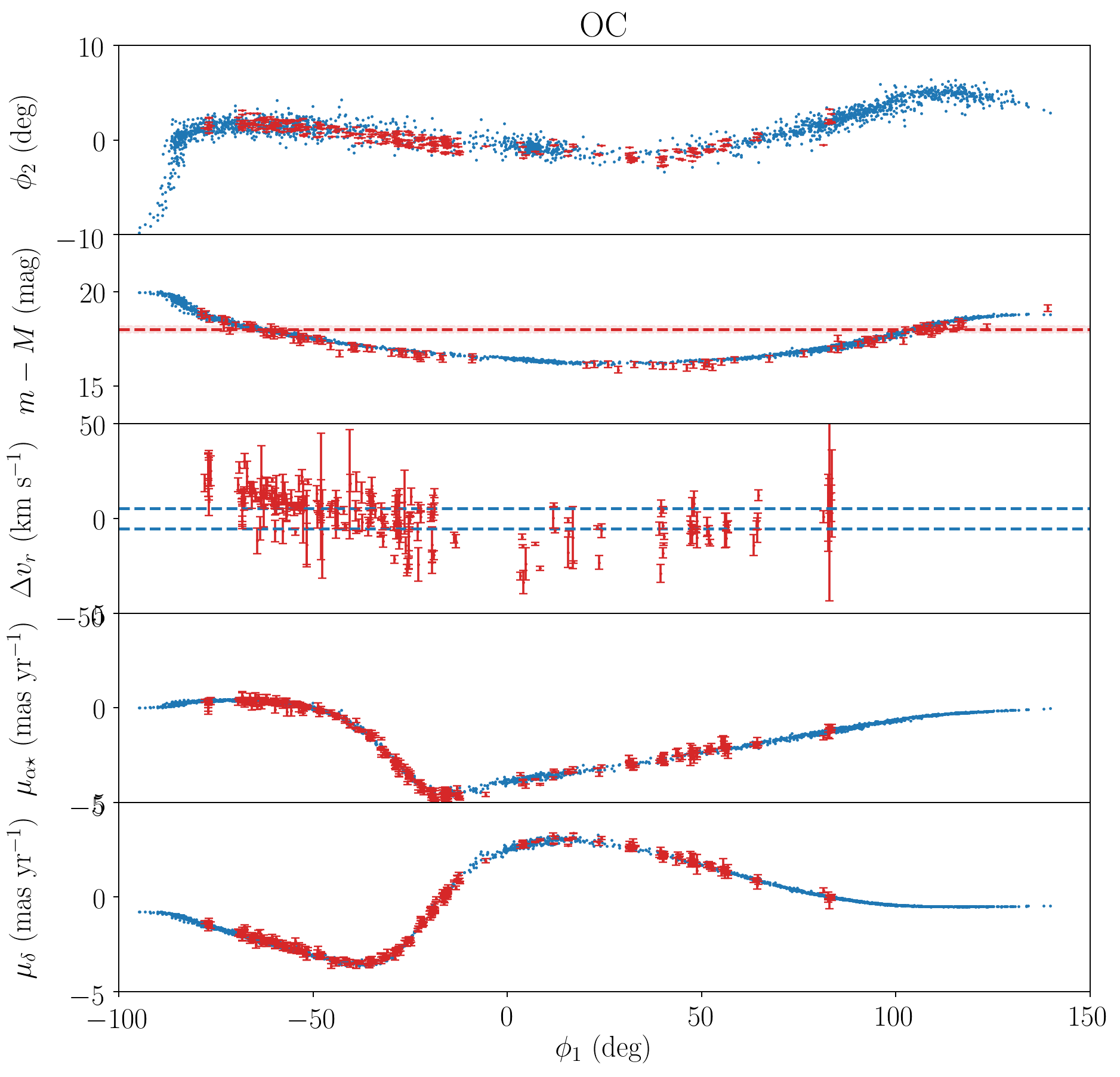}{0.45\textwidth}{(b)}}
    \gridline{\fig{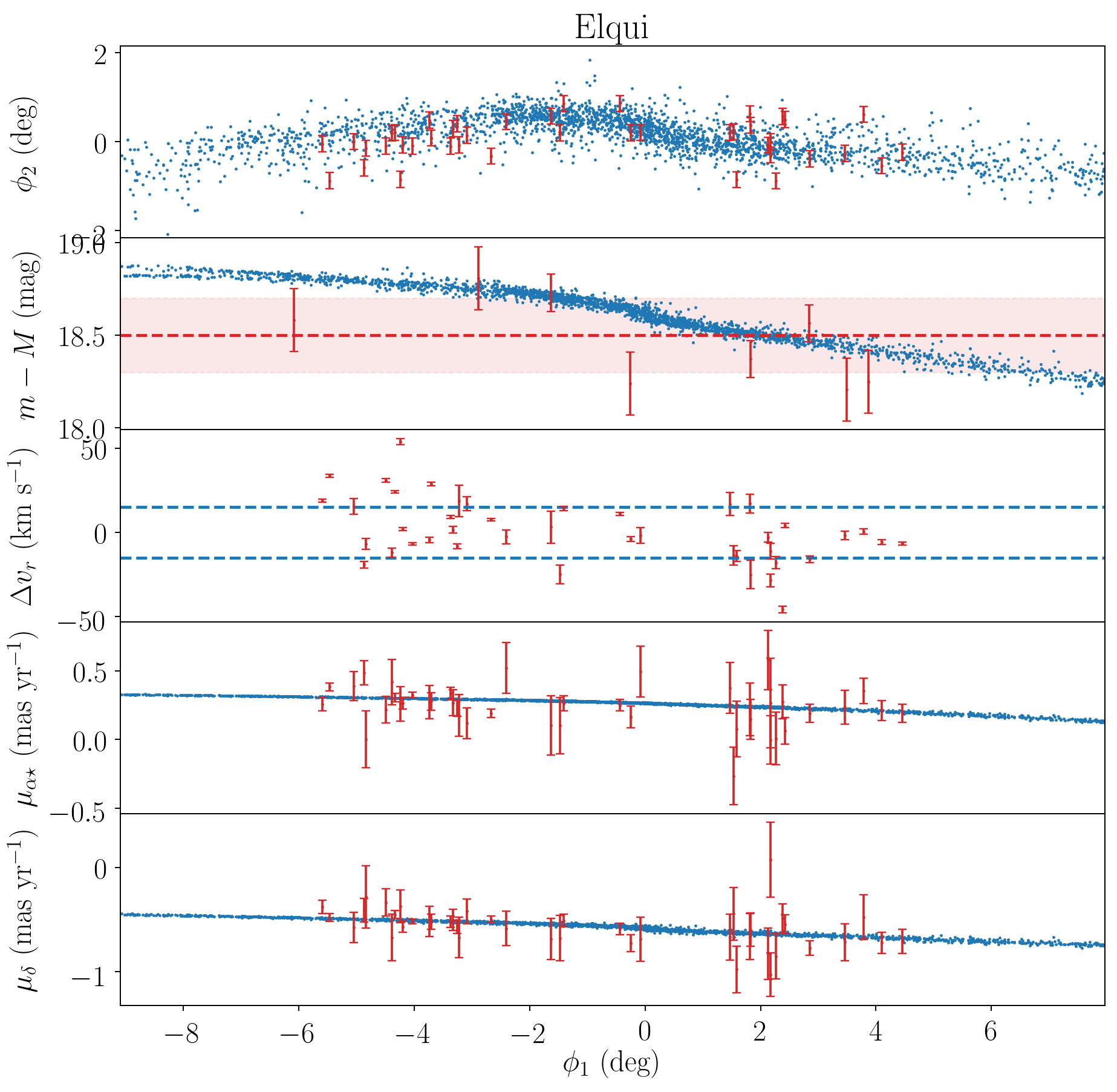}{0.45\textwidth}{(c)}
              \fig{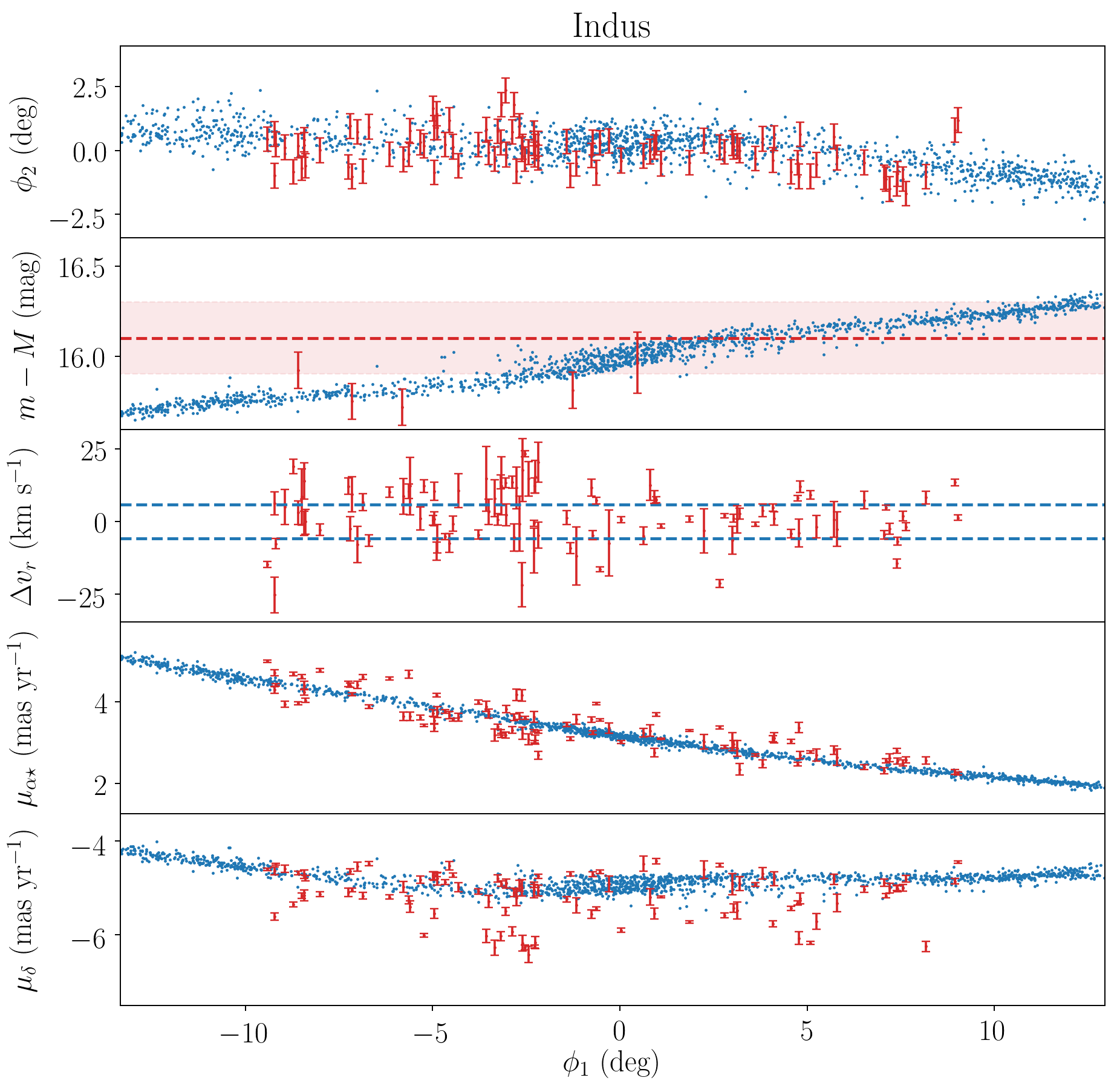}{0.45\textwidth}{(d)}}
    \caption{Stream models for (a) ATLAS, (b) OC, (c) Elqui, and (d) Indus.}
    \label{fig:models}
\end{figure}
\clearpage
\begin{figure}[htp]
    \centering
    \gridline{\fig{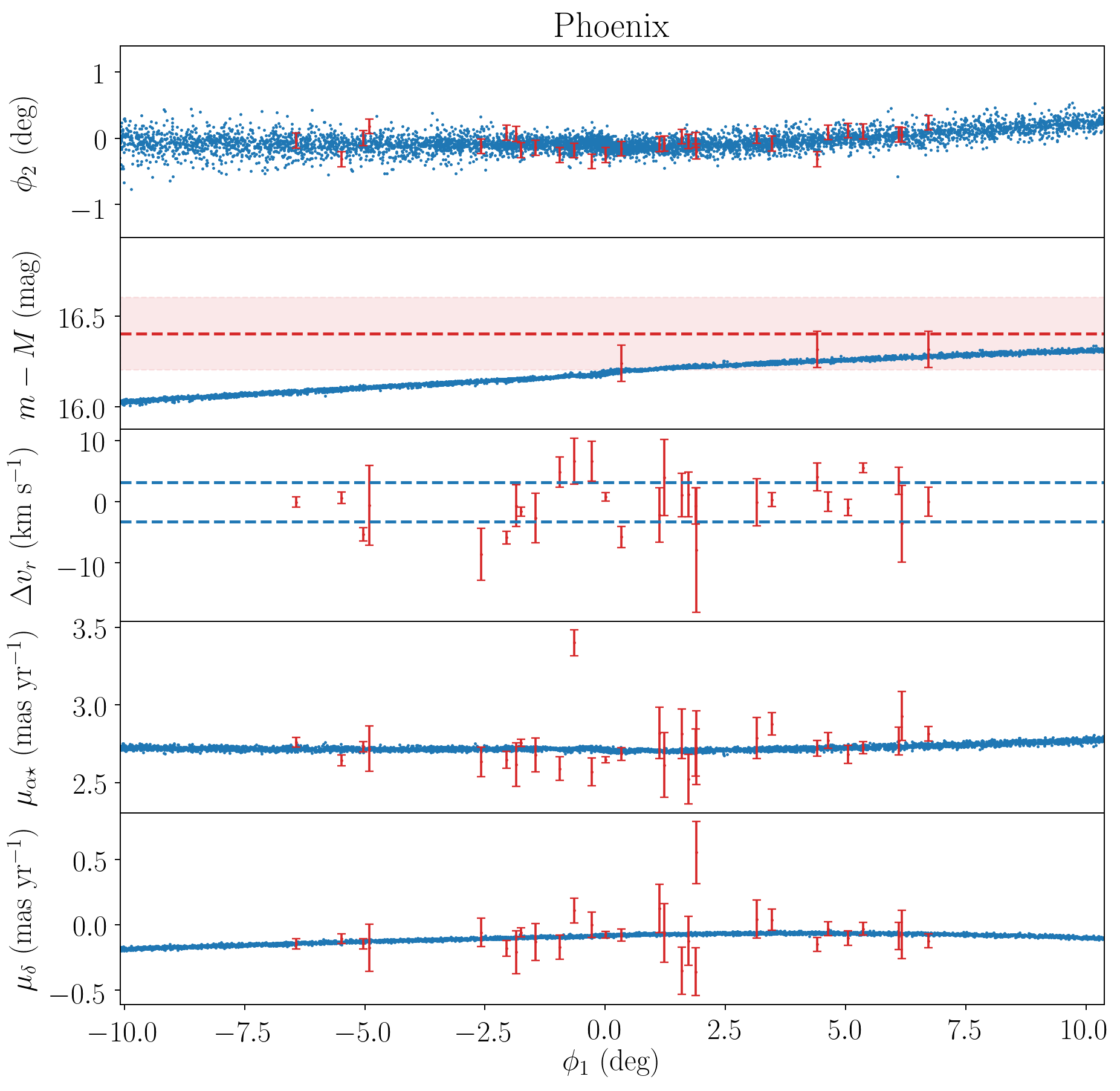}{0.45\textwidth}{(e)}
              \fig{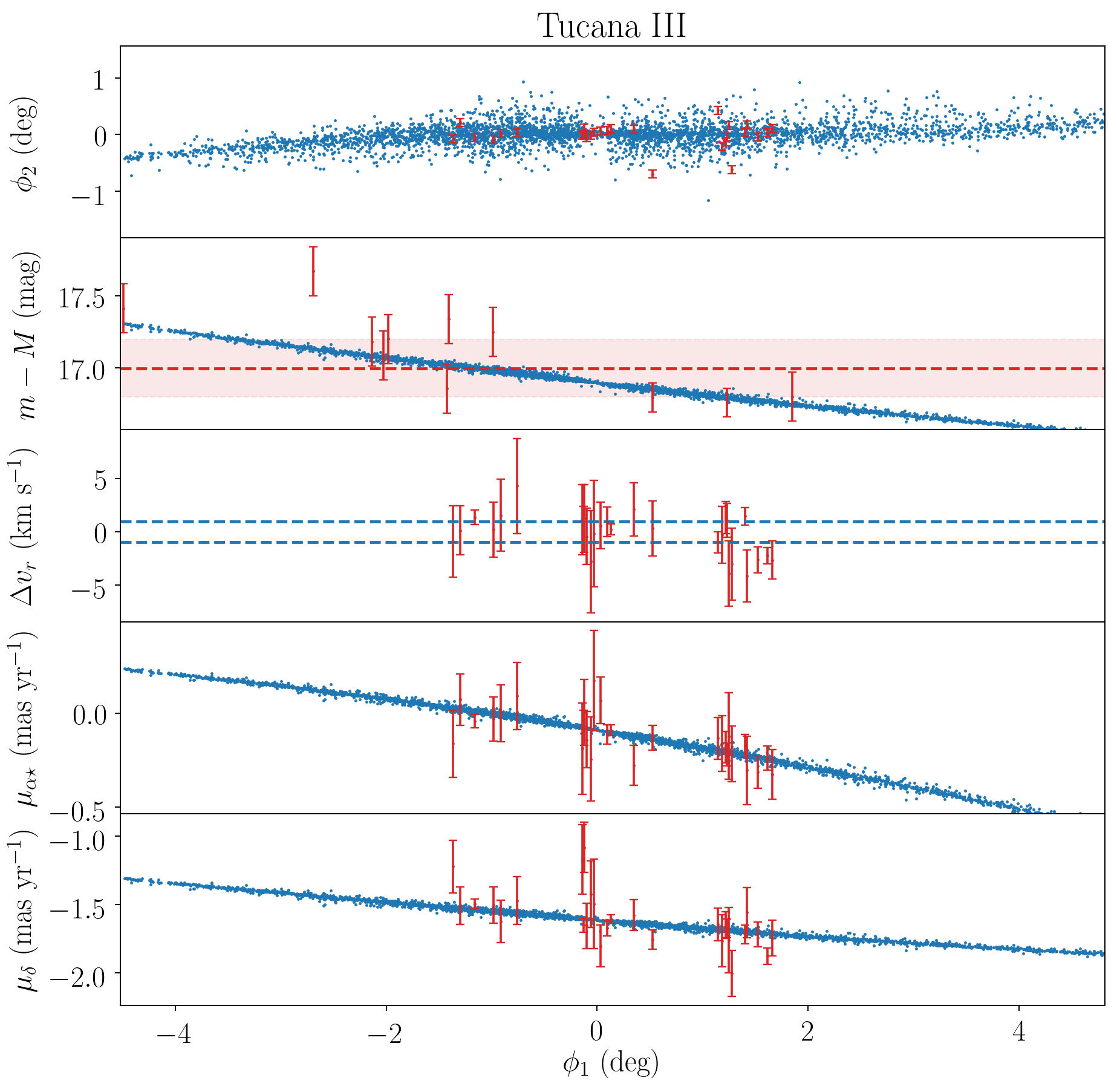}{0.45\textwidth}{(f)}}
    \gridline{\fig{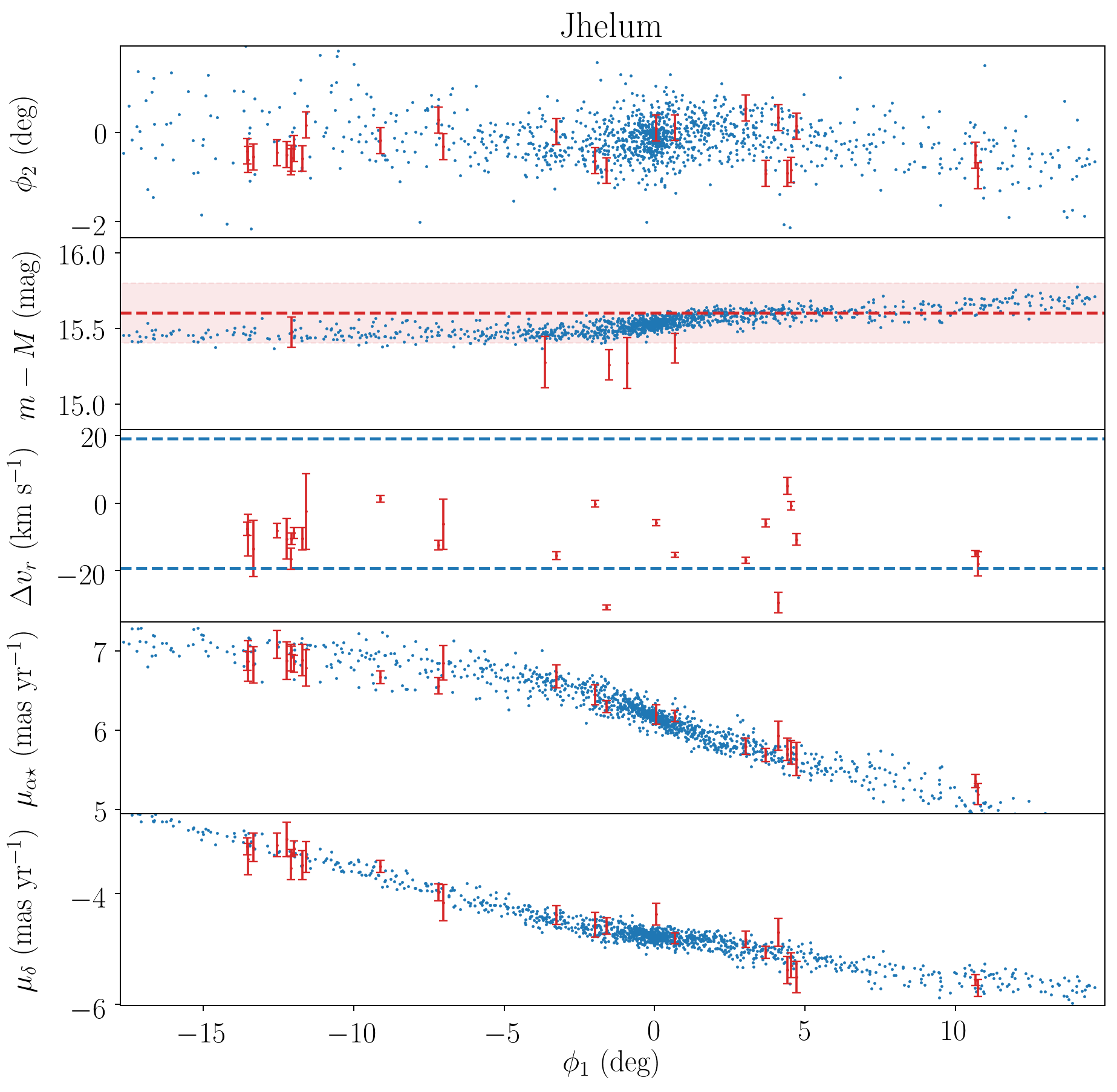}{0.45\textwidth}{(g)}}
    \caption{Stream models for (e) Phoenix, (f) Tucana III, and (g) Jhelum.}
    \label{fig:models2}
\end{figure}

\end{document}